\documentclass[aps,prl,onecolumn,preprint,showpacs,superscriptaddress,groupedaddress]{revtex4-1}
\usepackage{amsmath,amssymb}
\usepackage{graphicx}
\usepackage{times,hhline}
\usepackage{amsmath,bm}
\usepackage{epstopdf}
\newcommand{\sech}{\mbox{sech}}
\begin{document}
\title{Nondegenerate solitons in Manakov system}
\author{S. Stalin}
\affiliation{Centre for Nonlinear Dynamics, School of Physics, Bharathidasan University, Tiruchirapalli--620 024, India\\}
\author{R. Ramakrishnan}
\affiliation{Centre for Nonlinear Dynamics, School of Physics, Bharathidasan University, Tiruchirapalli--620 024, India\\}
\author{M. Senthilvelan }
\affiliation{Centre for Nonlinear Dynamics, School of Physics, Bharathidasan University, Tiruchirapalli--620 024, India\\}
\author{M. Lakshmanan \footnote{Corresponding author E-mail: lakshman@cnld.bdu.ac.in}}
\affiliation{Centre for Nonlinear Dynamics, School of Physics, Bharathidasan University, Tiruchirapalli--620 024, India\\} 
\begin{abstract}
It is known that Manakov equation which describes wave propagation in two mode optical fibers, photorefractive materials, etc. can admit solitons which allow energy redistribution between the modes on collision that also leads to logical computing. In this paper, we point out that Manakov system can admit more general type of nondegenerate fundamental solitons corresponding to different wave numbers, which undergo collisions without any energy redistribution. The  previously known class of solitons which allows energy redistribution among the modes turns out to be a special case corresponding to solitary waves with identical wave numbers in both the modes and travelling with the same velocity. We trace out the reason behind such a possibility and analyze the physical consequences.  
\end{abstract}
\maketitle
Discovery of solitons has created a new pathway to understand the wave propagation in many physical systems with nonlinearity \cite{a}. In particular, the existence  of optical solitons in nonlinear Kerr media \cite{b} provoked the investigation on solitons from different perspectives, particularly from applications point of view. By generalizing the waves propagating in an isotropic medium \cite{c} to an anisotropic medium, a pair of coupled equations for orthogonally polarized waves has been obtained by Manakov \cite{d,d1} as
\begin{equation}
iq_{jz}+q_{jtt}+2\sum_{p=1}^{2}|q_{p}|^{2}q_j=0,~~~~j=1,2
\label{e1}
\end{equation}
where $q_{j}$, $j=1,2$,  describe orthogonally polarized complex waves. Here the subscripts $z$ and $t$ represent normalized distance and retarded time, respectively. Equation (\ref{e1}) also appears in many physical situations such as single optical field propagation in birefringent fibers \cite{e}, self trapped incoherent light beam propagation in photorefractive medium \cite{f,g,h} and so on. Generalization of Eq. (\ref{e1}) to arbitrary $N$-waves is useful to model optical pulse propagation in multi-mode fibers \cite{i}. It has been identified \cite{d} that the polarization vectors of the solitons change when orthogonally polarized waves nonlinearly interact with each other leading to energy exchange interaction between the modes \cite{j}. Experimental observation of the latter has been demonstrated in \cite{k,l,l11}. The shape changing collision property of such waves, which we designate here as degenerate polarized soliton propagating with identical velocity and wave number in the two modes, gave rise to the possibility of constructing logic gates leading to all optical computing atleast in a theoretical sense \cite{l1,m,ma}. Energy sharing collisions among the optical vector solitons has been explored \cite{m} by constructing multi-soliton solutions explicitly to the multi-component nonlinear Schr\"{o}dinger equations. Further, it has been shown that the multi-soliton interaction process satisfies Yang-Baxter relation \cite{m1}. It is clear from these studies that the shape changing collision that occurs among the solitons with identical wave numbers in all the modes has been well understood. However, to our knowledge,  studies on solitons with non-identical wave numbers in all the modes have  not been considered so far. Consequently one would like to explore the role of such additional wave number(s) on the soliton structures and collision scenario as well.

In the contemporary studies, a new class of multi-hump solitons has been identified in different physical situations. In birefringent dispersive nonlinear media, asymmetric double hump-single hump frozen states have been obtained \cite{n}. Double hump structure has been observed for the Manakov equation by considering two soliton solutions \cite{o,p}. The first experimental observation of multi-hump solitons was demonstrated when the self trapped incoherent wave packets propagate in a dispersive nonlinear medium \cite{q}. These unusual solitons have been found in various nonlinear coupled field models \cite{t}. Stability of multi-hump optical solitons has also been investigated in the case of saturable nonlinear medium \cite{u}. It is reported that in such a medium both two and three hump solitons do not survive after collision. $N$-self trapped multi-humped partially coherent solitons have also been explored in photorefractive medium \cite{v}. The coherent coupling between copropagating fields also give rise to double hump solitons in the coherently coupled nonlinear Schr\"{o}dinger system \cite{w1}.  In addition to the above, the dynamics of multi-hump structured solitons have also been studied in certain dissipative systems \cite{new1,new2,new2a,new2b}. A double hump phase-locked higher-order vector soliton has been observed and its dynamics has been investigated in mode-locked fiber lasers \cite{new1,new2}.  Similarly in deployed fiber systems and fiber laser cavities, double hump solitons have been observed during the buildup process of soliton molecules \cite{new3,new4}.  

Motivated by the above, in this letter we present a new class of generalized soliton solutions for the Manakov model, exhibiting  various interesting structures under general parametric conditions. A fundamental double hump soliton (as well as other structures described below) sustains its shape even after collision with another similar soliton. This  behaviour is in contrast to the one which exists in saturable nonlinear media where two and three humps do not survive after collision. The soliton solutions presented in this letter also have both symmetric and asymmetric natures analogous to the partially coherent solitons in photorefractive medium. Under specific parametric restriction on wave numbers they degenerate into the standard Manakov solitons exhibiting shape changing collisions \cite{j,m}.

To explore the new family of soliton solutions for Eq. (\ref{e1}), we consider the bilinear forms of Eq. (\ref{e1}) as $(iD_z+D^2_t)g^{(j)} \cdot f=0$, $j=1,2$, and $D^2_t f \cdot f=2 \sum_{n=1}^{2}g^{(n)}g^{(n)*}$, which are obtained through the dependent variable transformations, $q_{j}=g^{(j)}/f$, $j=1,2$. Here $D_{z}$ and $D_{t}$ are the well known Hirota bilinear operators \cite{wa}, $g^{(j)}(z,t)$ are complex functions whereas $f$ is a real function and $*$ denotes complex conjugation. In principle, multi-soliton solutions of Eq. (\ref{e1}) can be constructed by solving recursively the system of linear partial differential equations which results in by substituting the series expansions $g^{(j)}=\epsilon g_1^{(j)}+\epsilon^3 g_3^{(j)}+...$ and $f=1+\epsilon^2 f_2+\epsilon^4 f_4+...$ for the unknown functions $g^{(j)}$ and $f$ in the bilinear forms. Here $\epsilon$ is a formal expansion parameter. 

 Considering two different seed solutions for $g_{1}^{(1)}$ and $g_{1}^{(2)}$ as $\alpha_{1}^{(1)}e^{\eta_1}$ and  $\alpha_{1}^{(2)}e^{\xi_1}$, respectively, where $\eta_{1}=k_{1}t+ik_{1}^{2}z$, $\xi_{1}=l_{1}t+il_{1}^{2}z$, and $\alpha_{1}^{(j)}$, $j=1,2$, $k_{1}$ and $l_{1}$ are in general independent complex wave numbers, to the resultant linear partial differential equations $(iD_z+D^2_t)g^{(j)}_1 \cdot 1=0$, $j=1,2$, which arise in the lowest order of $\epsilon$, the series expansion gets terminated as $g^{(j)}=\epsilon g_1^{(j)}+\epsilon^3 g_3^{(j)}$~~and~~ $f=1+\epsilon^2 f_2+\epsilon^4 f_4$. The explicit forms of the unknown functions present in the truncated series expansions constitute a new fundamental one soliton solution to Eq. (\ref{e1}) in the form
\begin{eqnarray}
q_{1}=(\alpha_{1}^{(1)} e^{\eta_{1}}+e^{\eta_{1}+\xi_{1}+\xi_{1}^*+\Delta_{1}^{(1)}})/D_1 \nonumber\\ 
q_{2}=(\alpha_{1}^{(2)} e^{\xi_{1}}+e^{\eta_{1}+\eta_{1}^*+\xi_{1}+\Delta_{1}^{(2)}})/D_1,
\label{e2}
\end{eqnarray}
where $D_1=1+e^{\eta_{1}+\eta_{1}^{*}+\delta_{1}}+e^{\xi_{1}+\xi_{1}^{*}+\delta_{2}}+e^{\eta_{1}+\eta_{1}^{*}+\xi_{1}+\xi_{1}^{*}+\delta_{11}}$, $e^{\delta_{1}}=\frac{|\alpha_{1}^{(1)}|^2}{(k_{1}+k_{1}^{*})^{2}}$, $e^{\delta_{2}}=\frac{ |\alpha_{1}^{(2)}|^2}{(l_{1}+l_{1}^*)^{2}}$, 
$e^{\delta_{11}}=\frac{|k_{1}-l_{1}|^{2} |\alpha_{1}^{(1)}|^{2}|\alpha_{1}^{(2)}|^{2}}{(k_{1}+k_{1}^*)^{2}(k_{1}^*+l_{1})(k_{1}+l_{1}^*)(l_{1}+l_{1}^*)^{2}}$, $e^{\Delta_{1}^{(1)}}=\frac{(k_{1}-l_{1})\alpha_{1}^{(1)}|\alpha_{1}^{(2)}|^2}{(k_{1}+l_{1}^*)(l_{1}+l_{1}^*)^{2}}$,
$e^{\Delta_{1}^{(2)}}=-\frac{(k_{1}-l_{1})|\alpha_{1}^{(1)}|^2\alpha_{1}^{(2)}}{(k_{1}+k_{1}^*)^{2}(k_{1}^*+l_{1})}$. 
\begin{figure}
\centering
\includegraphics[width=0.45\linewidth]{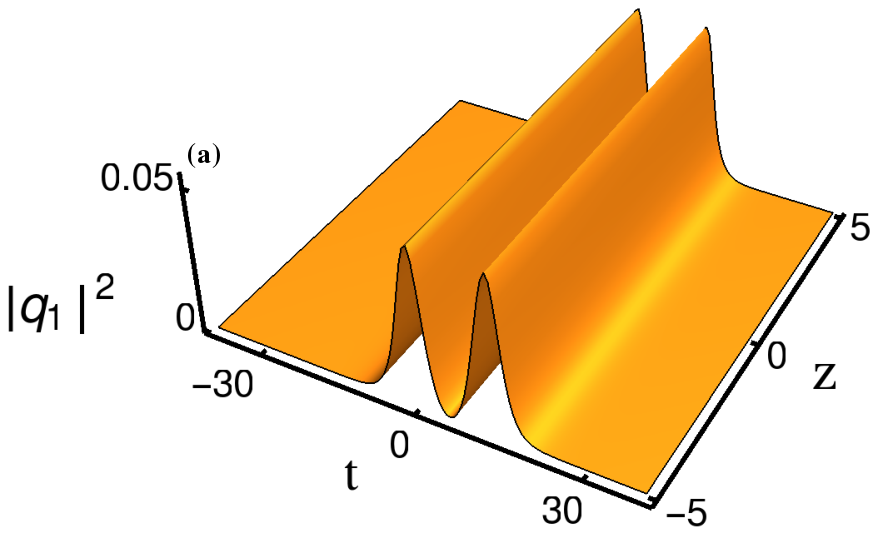}~~
\includegraphics[width=0.45\linewidth]{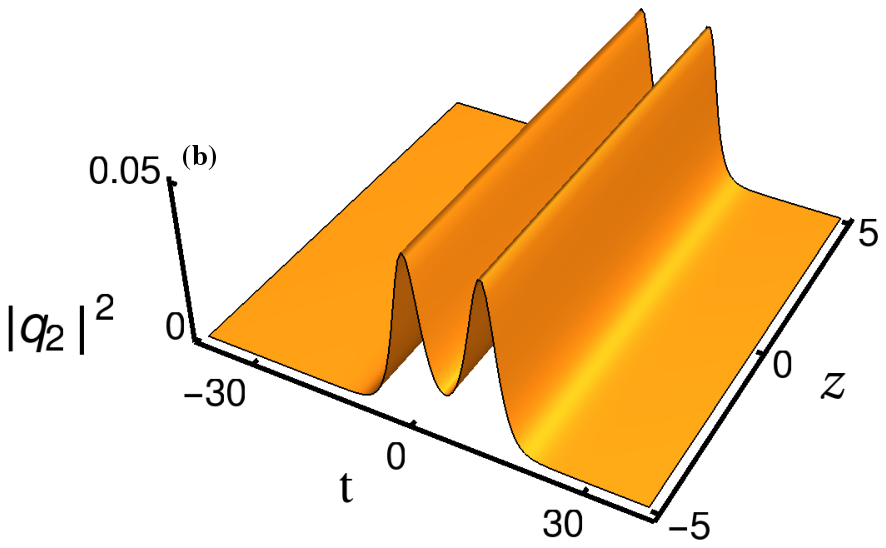}
\caption{Nondegenerate symmetric double hump one-soliton in the two modes: (a) and (b) denote the intensities of the components $q_{1}$ and $q_{2}$, respectively.   The parameters are chosen as   $k_{1}=0.316+0.5i$, $l_{1}=0.333+0.5i$, $\alpha_{1}^{(1)}=0.49+0.45i$ and $\alpha_{1}^{(2)}=0.45+0.45i$. }
\label{f1}
\end{figure}
 From the above, it is evident that the fundamental solitons propagating in the two modes are characterized by four arbitrary complex parameters $k_{1},l_{1}$ and $\alpha_{1}^{(j)}, j=1,2$. These nontrivial parameters determine the shape, amplitude, width and velocity of the solitons which propagate in the Kerr media or photorefractive media. The amplitudes of the solitons that are present in the two modes $q_{1}$ and $q_{2}$ are governed by the real parts of the wave numbers $k_{1}$ and $l_{1}$ whereas velocities are described by the imaginary parts of them. Note that $\alpha_1^{(j)}$, $j=1,2$, are related to  the unit polarization vectors of the solitons in the two modes. 

 To identify certain special features of the obtained four complex parameter family of soliton solution (\ref{e2}) we first consider  (for simplicity of analysis) the special case where the imaginary parts of the wave numbers $k_{1I}=l_{1I}$ but with $k_{1R}\neq l_{1R}$. The latter case yields atleast the following four different symmetric wave profiles, apart from similar asymmetric wave profiles, from solution (\ref{e2}) by incorporating the condition $k_{1R}< l_{1R}$ with further conditions and with suitable choices of parameters (examples given in \cite{w2}):  (i) single hump-single hump soliton: $\alpha_{1R}^{(1)}>\alpha_{1R}^{(2)}$ and $\alpha_{1I}^{(1)}=\alpha_{1I}^{(2)}$, (ii) double hump-single hump soliton: $\alpha_{1R}^{(1)}=\alpha_{1R}^{(2)}$ and $\alpha_{1I}^{(1)}<\alpha_{1I}^{(2)}$, (iii) double hump-flat top soliton: $\alpha_{1R}^{(1)}=\alpha_{1R}^{(2)}$ and $\alpha_{1I}^{(1)}\approx\alpha_{1I}^{(2)}$, (iv) double hump-double hump soliton:  $\alpha_{1R}^{(1)}>\alpha_{1R}^{(2)}$ and $\alpha_{1I}^{(1)}=\alpha_{1I}^{(2)}$. Similar conditions can be given for $k_{1R}> l_{1R}$ also. We have  not listed the asymmetric wave profiles here for brevity, which also exhibit the properties discussed below. Similar classification can be made for the case $k_{1I}\neq l_{1I}$, so that the solitons propagate in the two modes with different velocities, and exhibit similar interaction properties. These will be discussed separately.

To illustrate the symmetric case, we display only the intensity profile of the double hump soliton in Fig. \ref{f1}. We call the solitons that have two distinct wave numbers in both the  modes as in Eq. (\ref{e2}) as nondegenerate solitons (which can exist as different profiles as described above) while the solitons which have identical wave numbers in all the modes (which exist only  in single hump form) are designated as degenerate solitons. In particular, in the special case when $k_{1}=l_{1}$, the forms of $q_j$ given in Eq. (\ref{e2}) degenerate into the standard bright soliton form \cite{d,j}
\begin{equation}
q_{j}=\frac{\alpha_{1}^{(j)}e^{\eta_1}}{1+e^{\eta_1+\eta_1^*+R}}, ~j=1,2,\label{e4}\\
\end{equation}
which can be rewritten as
\begin{equation}
q_{j}=k_{1R} \hat{A_{j}}e^{i\eta_{1I}}\sech(\eta_{1R}+\frac{R}{2}),\label{e5}
\end{equation}
where $\eta_{1R}=k_{1R}(t-2k_{1I}z)$, $\eta_{1I}=k_{1I}t+(k_{1R}^{2}-k_{1I}^{2})z$, $\hat{A_{j}}=\frac{\alpha_{1}^{(j)}}{\sqrt{(|\alpha_{1}^{(1)}|^2+|\alpha_{1}^{(2)}|^2)}}$, $e^{R}=\frac{(|\alpha_{1}^{(1)}|^2+|\alpha_{1}^{(2)}|^2)}{(k_{1}+k_{1}^*)^2}$, $j=1,2$. 
Note that the above fundamental bright soliton always propagates in both the modes $q_{1}$ and $q_{2}$ with the same velocity $2k_{1I}$. The polarization vectors $(\hat{A}_1,\hat{A}_2)^{\dagger}$ have different amplitudes and phases, unlike the case of nondegenerate case where they  have only different phases $(A_1=(\frac{\alpha_{1}^{(1)}}{\alpha_{1}^{(1)*}})^{\frac{1}{2}},A_2=(\frac{\alpha_{1}^{(2)}}{\alpha_{1}^{(2)*}})^{\frac{1}{2}})^{\dagger}$ (vide Eq. (\ref{e2})), but same unit amplitude.  We call  the above type of soliton (\ref{e4}) or (\ref{e5}) as degenerate soliton \cite{w11}. 
\begin{figure}
\centering
\includegraphics[width=0.45\linewidth]{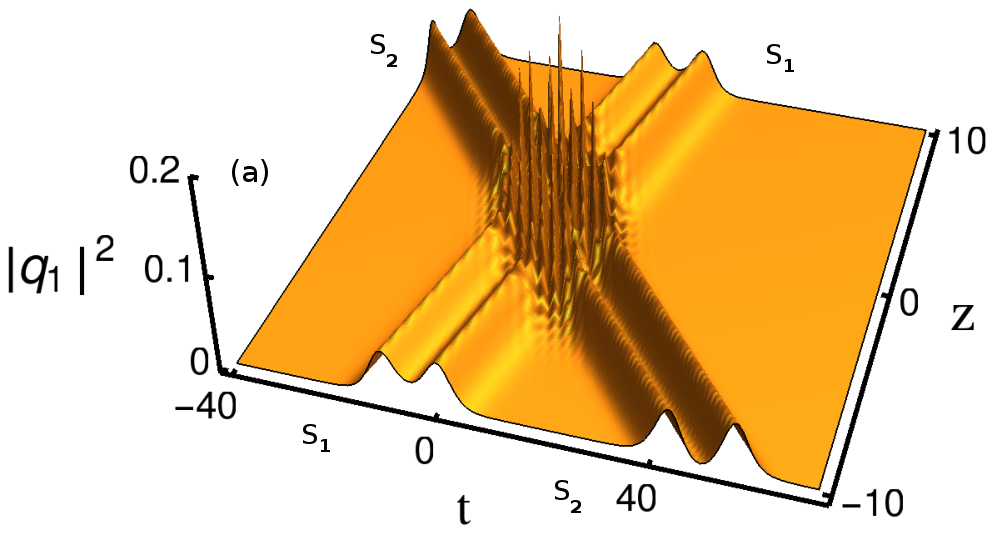}
\includegraphics[width=0.45\linewidth]{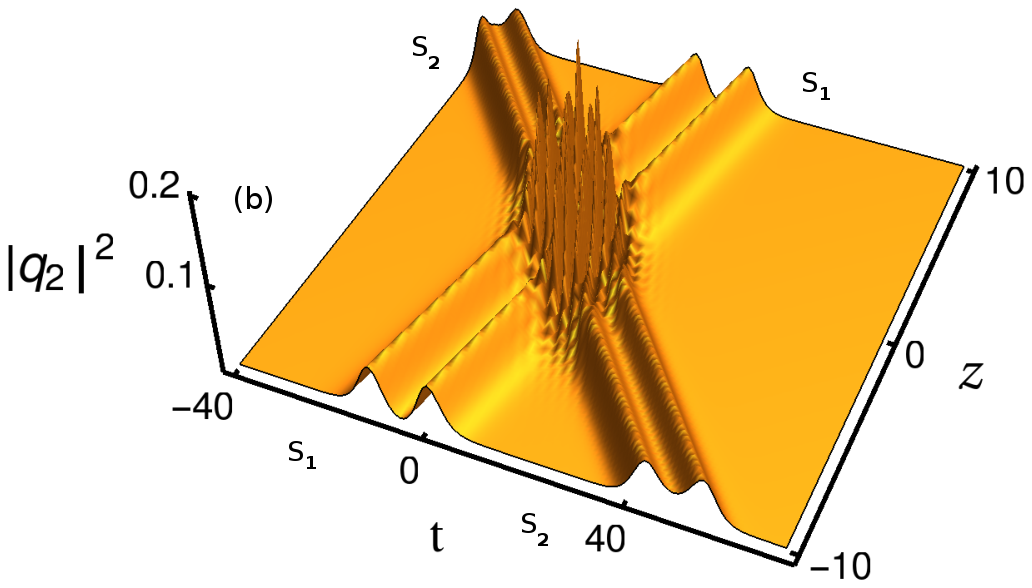}
\caption{Nondegenerate solitons exhibiting shape preserving collisions: (a) and (b) denote the elastic collision of two symmetric double hump solitons for the parametric values $k_{1} = 0.333 + 0.5i$, $k_{2} = 0.3 - 2.2i$, $l_{1} = 0.3 + 0.5i$, $l_{2} = 0.333 - 2.2i$, $\alpha_{1}^{(1)} = 0.45 + 0.45i$, $\alpha_{2}^{(1)} =0.49 + 0.45i$,  $\alpha_{1}^{(2)} = 0.49 + 0.45i$ and $\alpha_{2}^{(2)} = 0.45 + 0.45i$.}
\label{f2}
\end{figure}

 In order to understand the collision dynamics of the soliton solution of the kind (\ref{e2}), it is essential to construct the corresponding two soliton solution. In the latter case, the series expansion for $q_j$, $j=1,2$, gets terminated as $g^{(j)}=\epsilon g_1^{(j)}+\epsilon^3 g_3^{(j)}+\epsilon^5 g_5^{(j)}+\epsilon^7 g_7^{(j)}$ and $f=1+\epsilon^2 f_2+\epsilon^4 f_4+\epsilon^6 f_6+\epsilon^8 f_8$. The obtained explicit forms of $g^{(j)}$ and $f$, $j=1,2$, in the above truncated expansions constitute the nondegenerate two soliton solution of Eq. (\ref{e1}), which reduces to the known form given in Ref. \cite{j} for $k_i=l_i$, $i=1,2$. The complicated profiles of the present nondegenerate two soliton solution are governed by eight arbitrary complex parameters $k_{j}$, $l_{j}$, $\alpha_{1}^{(j)}$ and $\alpha_{2}^{(j)}$, $j=1,2$ (see supplemental material \cite{w2}). 

 To study the collision dynamics between the nondegenerate two solitons, as an example, we again confine ourselves to the case of symmetric double hump solitons  by fixing the imaginary parts of the wave numbers as $k_{iI}=l_{iI}$, $i=1,2$.  For other types also similar analysis has been carried out. By carefully examining the behavior of the obtained nondegenerate two soliton solution in the asymptotic regimes, $z \rightarrow \pm \infty$, we find that the phases of the fundamental nondegenerate double hump solitons in both the modes change during the collision process, while the intensities remain unchanged. This can be verified by defining the transition amplitudes as $T_{j}^{l}$ $=$ $\frac{A_{j}^{l+}}{A_{j}^{l-}}$~,~$j$ $=$1,2 and $l=1,2$, where subscript $j$ represents the mode and superscript $l\pm$ denote the nondegenerate soliton numbers 1 and 2 designated as $S_{1}$ and $S_{2}$, respectively, in the asymptotic regimes $z \rightarrow \pm \infty$.  

\begin{figure}
\centering
\includegraphics[width=0.45\linewidth]{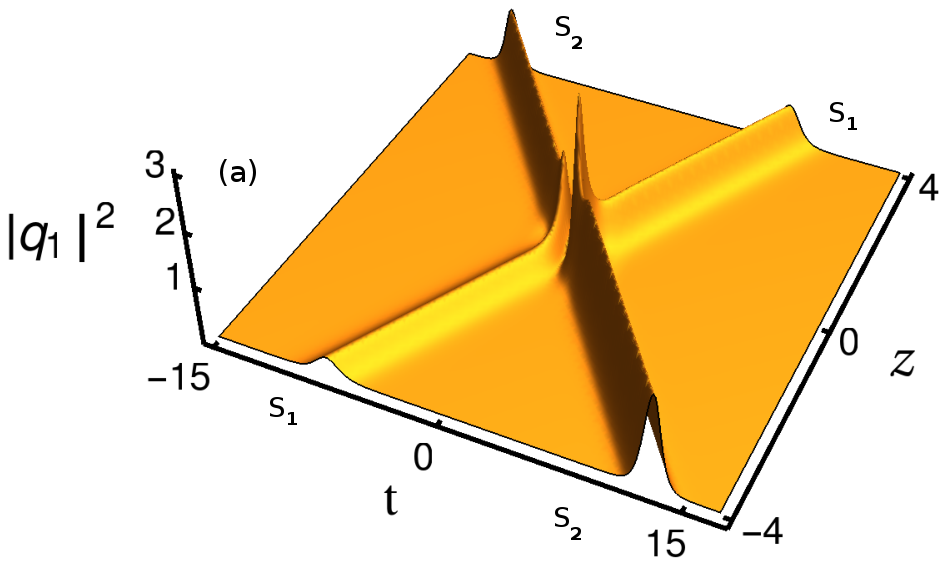}
\includegraphics[width=0.45\linewidth]{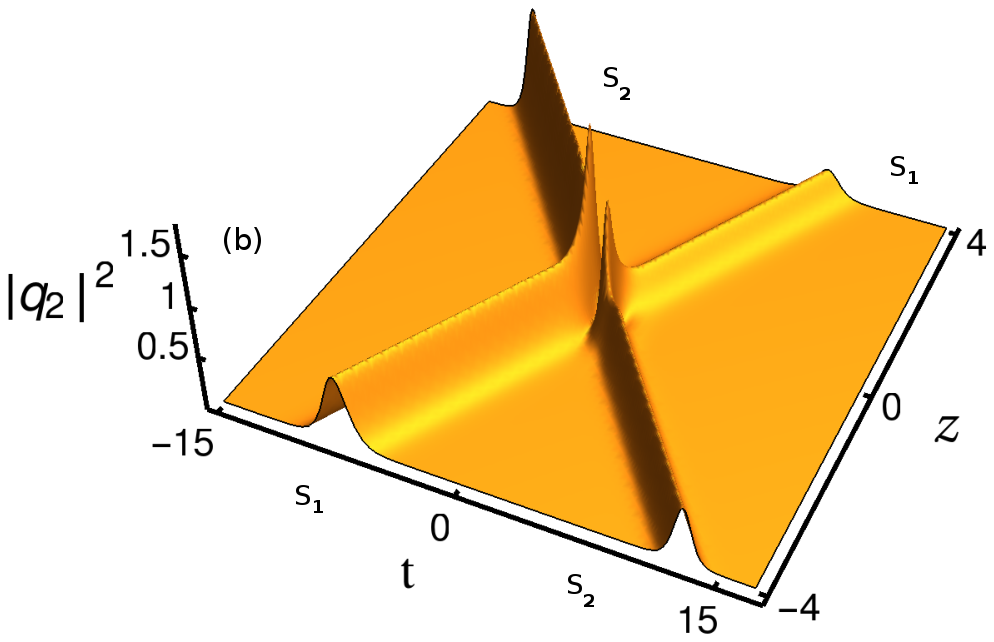}
\caption{Degenerate solitons exhibiting shape changing collision: (a) and (b) denote the energy sharing collision in the two modes for the parametric values $k_{1} =l_1= 1+ i$, $k_{2} = l_2=1.51-1.5i$, $\alpha_{1}^{(1)} =0.5 +0.5 i$, $\alpha_{2}^{(1)} =\alpha_{1}^{(2)}= \alpha_{2}^{(2)}= 1$.}
\label{f3}
\end{figure}
In the nondegenerate double hump soliton case, the amplitudes of the solitons $S_{1}$ and $S_{2}$ in the first mode $\big(2k_{1R}A_{1}^{1-}$, $\frac{(k_{1}-k_{2})(k_{2}-l_{1})^{1/2}(k_{1}+k_{2}^*)(k_{2}^*+l_{1})^{1/2}}{(k_{1}^*-k_{2}^*)(k_{2}^*-l_{1}^*)^{1/2}(k_{1}^*+k_{2})(k_{2}+l_{1}^*)^{1/2}}2k_{2R}A_{2}^{1-}\big)$ before collision change to $\big(\frac{(k_{1}-k_{2})(k_{1}-l_{2})^{1/2}(k_{1}^*+k_{2})(k_{1}^*+l_{2})^{1/2}}{(k_{1}^*-k_{2}^*)(k_{1}^*-l_{2}^*)^{1/2}(k_{1}+k_{2}^*)(k_{1}+l_{2}^*)^{1/2}}2k_{1R}A_{1}^{1+}$, $2k_{2R}A_{2}^{1+}\big)$ after collision, where $A_{1}^{1\pm}=\sqrt{\frac{\alpha_{1}^{(1)}}{\alpha_{1}^{(1)^*}}}$, $A_{2}^{1\pm}=\sqrt{\frac{\alpha_{2}^{(1)}}{\alpha_{2}^{(1)^*}}}$. Similarly in the second component the amplitudes of the solitons $S_{1}$ and $S_{2}$ are  $\big(2l_{1R}A_{1}^{2-}$, $\frac{(l_{1}-l_{2})(k_{1}-l_{2})^{1/2}(l_{1}+l_{2}^*)(k_{1}+l_{2}^*)^{1/2}}{(l_{1}^*-l_{2}^*)(k_{1}^*-l_{2}^*)^{1/2} (k_{1}^*+l_{2})^{1/2}(l_{1}^*+l_{2})}2l_{2R}A_{2}^{2-}\big)$ before collision which change to $\big(\frac{(1_{1}-l_{2})(l_{1}-k_{2})^{1/2}(l_{1}^*+l_{2})(k_{2}+l_{1}^*)^{1/2}}{(l_{1}^*-l_{2}^*)(k_{2}^*-l_{1}^*)^{1/2}(l_{1}+l_{2}^*)(k_{2}^*+l_{2})^{1/2}}2l_{1R}A_{1}^{2+}$, $2l_{2R}A_{2}^{2+}\big)$ after collision, where $A_{1}^{2\pm}=\sqrt{\frac{\alpha_{1}^{(2)}}{\alpha_{1}^{(2)^*}}}$, $A_{2}^{2\pm}=\sqrt{\frac{\alpha_{2}^{(2)}}{\alpha_{2}^{(2)^*}}}$. However the intensity redistribution does not occur among the modes of the solitons, which can be confirmed by taking the absolute squares of the transition amplitudes which turn out to be unity, that is $|T_{j}^{l}|^2=1$. This shows that in the nondegenerate case, $k_i\neq l_i$, $i=1,2$, the polarization vectors do not contribute to intensity redistribution among the modes. Consequently the double hump solitons in each mode exhibit shape preserving collision corresponding to elastic nature. This is illustrated in Fig. \ref{f2} for the parameter values given there by actually plotting the two-soliton solution (given in supplemental material). From this figure it is easy to identify that the intensity/energy of the double hump solitons in the two modes propagate without change after collision with another double hump soliton except for a phase shift. Similar scenario exists generally for all other cases of $k_i\neq l_i$, $i=1,2$, the details of which will be published elsewhere.  We also find that the phases of the soliton $S_1$ in the two modes change from $\big(\frac{\Delta_1^{(1)}-\rho_1}{2}, \frac{\Delta_1^{(2)}-\rho_2}{2}\big)$ to $\big(\frac{\phi_2^{(1)}-\mu_8^{(1)}}{2}, \frac{\phi_2^{(2)}-\mu_8^{(2)}}{2}\big)$ during the collision process, while the phases of soliton $S_2$ change from $\big(\frac{\phi_1^{(1)}-\mu_1^{(1)}}{2}, \frac{\phi_1^{(2)}-\mu_1^{(2)}}{2}\big)$ to $\big(\frac{\Delta_8^{(1)}-\hat{\rho}_1}{2}, \frac{\Delta_8^{(2)}-\hat{\rho}_2}{2}\big)$ after collision. Here $\rho_j=\log\alpha_1^{(j)}$, $\hat{\rho}_j=\log\alpha_2^{(j)}$, $\Delta_{1}^{(j)}$, $\phi_1^{(j)}$, $\phi_2^{(j)}$, $\mu_1^{(j)}$, $\mu_8^{(j)}$, $\Delta_{8}^{(j)}$, $j=1,2$  are constants (see\cite{w2}). 

 In addition to the above, we have also observed similar shape preserving collision in the case of symmetric single hump soliton when it collides with another identical soliton. The flat top soliton also preserves its structure when it collides with a symmetric double hump soliton. However, while testing the stability property of a double hump soliton interacting with a  single hump soliton, we come across a slightly different collision scenario. During this interaction process,  the symmetric double hump soliton experiences a strong perturbation due to the collision with the symmetric single hump soliton. The result of their collision  is only reflected in a change in the shape of the symmetric double hump soliton into a slightly  asymmetric form, but without change in energy. However, the symmetric single hump soliton does not undergo any change (see \cite{w2}). 

 In contrast to the nondegenerate case, the nonlinear superposition of degenerate fundamental solitons ($k_i=l_i$, $i=1,2$) in the Manakov system exhibits interesting shape changing collision due to intensity redistribution among the modes as shown in Ref. \cite{j}. The intensity redistribution occurs in the degenerate case due to the arbitrary polarization vectors in the two modes getting mixed up, which is illustrated in Fig. \ref{f3} where the intensity redistribution occurs because of the enhancement or suppression of intensity in any one of the modes in either one of the degenerate solitons with corresponding suppression or enhancement of intensity of the same soliton, see Eq. (\ref{e5}) \cite{j}. To hold the energy conservation between the two modes the intensities of the two solitons $S_1$ and $S_{2}$ change appropriately. It is well known that the degenerate soliton/Manakov soliton (Eq. (3)) reported in Refs. [10,11]  in general exhibits shape changing collision through energy redistribution among the modes (except for the very special case $\frac{\alpha_1^{(1)}}{\alpha_2^{(1)}}=\frac{\alpha_1^{(2)}}{\alpha_2^{(2)}}$ \cite{i,m}, where elastic collision occurs).  We have also verified the elastic nature of double hump soliton collision using Crank–Nicolson method \cite{pm}. 

 We also further wish to point out that considering the notion dissipative solitons, they also exhibit elastic collision property. However, this collision scenario, for example in a fiber laser cavity, is entirely different from the one that occur in our present case. In the fiber laser cavity, during the collision between  the soliton pair (bound state/doublet) and single soliton state (singlet), the single soliton destroys the bound state but another pair is formed that moves away with the same velocity leaving one of the solitons of the previously moving pair in rest \cite{dis1,dis3}. During this collision scenario, the energy or momentum  is not conserved in the dissipative system (fiber laser cavity). To bring the above elastic collision, it is essential to set up the binding energy of solitons to be non-zero and the difference in velocities of the pair and the singlet is fixed and must be same before and after collision \cite{dis1,dis3}. Also no explicit analytical form of such dissipative soliton is available for direct analysis.

In principle one can construct the $N$-soliton solution of the nondegenerate type to the Manakov system by following the procedure given above. For the $N$-nondegenerate soliton, the power series expansion should be of the form $g^{(j)}=\sum_{n=1}^{2N-1}\epsilon^{2n-1}g_{2n-1}^{(j)}$ and $f=1+\sum_{n=1}^{2N}\epsilon^{2n}f_{2n}$.  The shape of the profile will be determined by the $4N$ complex parameters which are present in the $N$-soliton solution. The degenerate soliton solutions can be recovered from the nondegenerate $N$-soliton solutions by fixing the wave numbers as $k_{i}=l_{i}, i=1,2,...,N$. The symmetric profile of the multi-nondegenerate soliton can be obtained by fixing the imaginary parts as $k_{iI}=l_{iI}, i=1,2,...N$. We also point out that the symmetric and asymmetric cases of the nondegenerate soliton solution given in Eq. (\ref{e2}) can be compared with partially incoherent soliton in photorefractive medium \cite{v}. The profile of the PCS is determined by only three real parameters for $N=2$ as special case of the degenerate soliton \cite{i,m} (Manakov case) whereas in the present nondegenerate case, the profiles of the single soliton are governed by four complex parameters. In the incoherent limit (number of modes is infinity) the shape of the PCS can be arbitrary since the number of parameters involved in the underlying analytical form is $N$-free real parameters. However in the incoherent limit the presence of $2N$ free complex parameters in the nondegenerate fundamental one soliton would bring  in more complex shapes than the above mentioned PCS reported in the photorefractive medium.  


 To  observe the existence of nondegenerate solitons (\ref{e2}) experimentally one may consider the mutual-incoherence procedure given in \cite{k,q} with two different laser sources of different characters (instead of a single laser source). Using polarizing beam splitters, the extraordinary beams coming out from the two laser sources can be further split into four individual incoherent fields. These four fields can act as nondegenerate two individual solitons in the two modes. Further, the collision angle must be large enough to to observe the elastic collision between these  nondegenerate two solitons in both the modes \cite{k,l}. The  experimental procedure with a single laser can be used to observe the Manakov solitons and multi-mode multi-hump solitons that arise in dispersive nonlinear medium \cite{k,q}.


Finally, it is essential to point out the application of our above reported soliton solutions. Our results open up a new possibility to investigate nondegenerate solitons in both integrable and non-integrable systems and will give rich coherent structures when the four wave mixing phenomenon is taken into account. Our studies can also be extended to fiber arrays and multi-mode fibers where the pulse propagation is described by Manakov type equations. Experimental observation of Manakov solitons in AlGaAs planar wave guides \cite{l} and multi-hump solitons in the multi-mode self-induced wave guides \cite{q} give the impression that our results will be important to an interaction of optical field in coupled field models.  The shape preserving collision which occurs among the nondegenerate solitons can be used  for the optical communication process. The double hump nature of the nondegenerate solitons can be useful for the information process as described in the concept of soliton molecule \cite{new3}. As far as the degenerate soliton is concerned, it has already been shown that it is useful in the the computation process \cite{l1,m}. We note that under appropriate conditions, namely $k_{1I}\approx k_{2I}$ and $l_{1I}\approx l_{2I}$,  the nondegenerate solitons reported in the present conservative system  can be seen as the soliton molecule observed in  the deployed fiber systems and in fiber laser cavities \cite{new3,new3a,new4,new4a,new4b}.

In conclusion, we have shown that the Manakov model under a general physical situation admits interesting nondegenerate solitons exhibiting shape preserving collisions thereby leading to explain the interaction of elastic nature of light-light interaction under general initial conditions. The fascinating energy sharing collisions exhibiting the nonlinear superposition of degenerate multi-solitons can be extracted from the nondegenerate soliton solutions under the specific physical restrictions  which leads to the construction of optical logic gates \cite{l1}.

The authors are thankful to Dr. T. Kanna for several valuable discussions on nondegenerate solitons in coupled nonlinear Schr\"{o}dinger equations. 
The work of R.R. and M. L. are supported by Department of Atomic Energy - National Board of Higher Mathematics (Grant no. 2/48(5)/2015/NBHM(R.P.)/R \& D II/14127). The work of M.S. is supported by the Department of Science and Technology (Grant No. EMR/2016/001818). The work of M.L.  is also supported by the Department of Science and Technology-Science and Engineering Research Board Distinguished Fellowship (Grant no. SERB/F/6717/2017-18). 

\end{document}


\title{Supplemental material: Nondegenerate solitons in Manakov system}
\author{S. Stalin}
\affiliation{Centre for Nonlinear Dynamics, School of Physics, Bharathidasan University, Tiruchirapalli--620 024, India\\}
\author{R. Ramakrishnan}
\affiliation{Centre for Nonlinear Dynamics, School of Physics, Bharathidasan University, Tiruchirapalli--620 024, India\\}
\author{M. Senthilvelan}
\affiliation{Centre for Nonlinear Dynamics, School of Physics, Bharathidasan University, Tiruchirapalli--620 024, India\\}
\author{M. Lakshmanan \footnote{Corresponding author  E-mail: lakshman@cnld.bdu.ac.in}}
\affiliation{Centre for Nonlinear Dynamics, School of Physics, Bharathidasan University, Tiruchirapalli--620 024, India\\} 
\maketitle
\section{S.1. Nondegenerate two-soliton solution}
To explore the nondegenerate two-soliton solution of Manakov system, we adopt the Hirota's bilinear method which we have discussed in the main text. We find that the Hirota's series expansion gets truncated for the nondegenerate two-soliton solution as
\begin{eqnarray}
&&g^{(j)}=\epsilon g_1^{(j)}+\epsilon^3 g_3^{(j)}+\epsilon^5 g_5^{(j)}+\epsilon^7 g_7^{(j)}, j=1,2,\nonumber\\
&&f=1+\epsilon^2 f_2+\epsilon^4 f_4+\epsilon^6 f_6+\epsilon^8 f_8.
\end{eqnarray}
The following nondegenerate two soliton solution of Manakov system can be obtained by finding the unknown functions that are present in the above series expansions,
\begin{equation*}
q_{j}=\frac{g^{(j)}}{f},~ j=1,2,
\end{equation*}                                                              
where the explicit forms of $g^{(j)}$ and $f$ are given by 
\begin{subequations}
\begin{eqnarray}
g^{(1)}&=&\sum_{j=1}^2 \alpha_{j}^{(1)}e^{\eta_j}+e^{\eta_1}\big(\sum_{i,j=1}^{2}e^{\xi_i+\xi_j^*+\Del_{ij}}+e^{\eta_1^*+\eta_2+\Del_{13}}+e^{\xi_1+\xi_1^*+\xi_2+\xi_2^*+\Del_{14}}\big)\nonumber\\
&&+e^{\eta_2}\big(\sum_{i,j=1}^{2}e^{\xi_i+\xi_j^*+\Lam_{ij}}+e^{\eta_1+\eta_2^*+\Del_{31}}+e^{\xi_1+\xi_1^*+\xi_2+\xi_2^*+\Del_{41}}\big)\nonumber\\
&&+e^{\eta_1+\eta_2+\eta_1^*}\big(\sum_{i,j=1}^{2}e^{\xi_i+\xi_j^*+\Th_{ij}}+e^{\xi_1+\xi_1^*+\xi_2+\xi_2^*+\Del_{15}}\big)\nonumber\\
&&+e^{\eta_1+\eta_2+\eta_2^*}\big(\sum_{i,j=1}^{2}e^{\xi_i+\xi_j^*+\Phi_{ij}}+e^{\xi_1+\xi_1^*+\xi_2+\xi_2^*+\Del_{51}}\big),\\
g^{(2)}&=&\sum_{j=1}^2 \alpha_{j}^{(2)}e^{\xi_j}+e^{\xi_1}\big(\sum_{i,j=1}^{2}e^{\eta_i+\eta_j^*+\ga_{ij}}+e^{\xi_1^*+\xi_2+\ga_{13}}+e^{\eta_1+\eta_1^*+\eta_2+\eta_2^*+\ga_{14}}\big)\nonumber\\
&&+e^{\xi_2}\big(\sum_{i,j=1}^{2}e^{\eta_i+\eta_j^*+\mu_{ij}}+e^{\xi_1+\xi_2^*+\ga_{31}}+e^{\eta_1+\eta_1^*+\eta_2+\eta_2^*+\ga_{41}}\big)\nonumber\\
&&+e^{\xi_1+\xi_2+\xi_1^*}\big(\sum_{i,j=1}^{2}e^{\eta_i+\eta_j^*+\nu_{ij}}+e^{\eta_1+\eta_1^*+\eta_2+\eta_2^*+\ga_{15}}\big)\nonumber\\
&&+e^{\xi_1+\xi_2+\xi_2^*}\big(\sum_{i,j=1}^{2}e^{\eta_i+\eta_j^*+\chi_{ij}}+e^{\eta_1+\eta_1^*+\eta_2+\eta_2^*+\ga_{51}}\big),
\end{eqnarray}
\begin{eqnarray}
f&=&1+e^{\eta_1+\eta_1^*}\big(e^{\del_1}+\sum_{i,j=1}^2e^{\xi_i+\xi_j^*+\del_{ij}}+e^{\xi_1+\xi_1^*+\xi_2+\xi_2^*+\del_{13}}\big)+e^{\eta_1^*+\eta_2}\big(e^{\del_2}\nonumber\\
&&+\sum_{i,j=1}^2e^{\xi_i+\xi_j^*+\vth_{ij}}+e^{\xi_1+\xi_1^*+\xi_2+\xi_2^*+\del_{14}}\big)+e^{\eta_1+\eta_2^*}\big(e^{\del_2^*}+\sum_{i,j=1}^2e^{\xi_i+\xi_j^*+\vphi_{ij}}\nonumber\\
&&+e^{\xi_1+\xi_1^*+\xi_2+\xi_2^*+\del_{15}}\big)+e^{\eta_2+\eta_2^*}\big(e^{\del_3}+\sum_{i,j=1}^2e^{\xi_i+\xi_j^*+\vsa_{ij}}+e^{\xi_1+\xi_1^*+\xi_2+\xi_2^*+\del_{16}}\big)\nonumber\\
&&+e^{\eta_1+\eta_1^*+\eta_2+\eta_2^*}\big(\sum_{i,j=1}^2e^{\xi_i+\xi_j^*+\phi_{ij}}+e^{\del_{17}}+e^{\xi_1+\xi_1^*+\xi_2+\xi_2^*+\del_{18}}\big)+\sum_{i,j=1}^2e^{\xi_i+\xi_j^*+\psi_{ij}}\nonumber\\
&&+e^{\xi_1+\xi_1^*+\xi_2+\xi_2^*+\del_{19}}
\end{eqnarray}
\end{subequations}
The constants that are present in the above two-soliton solution are given below,
\begin{eqnarray*}
&&e^{\Del_{11}}=\frac{\vrho_{11}\al_1^{(1)}\al_1^{(2)}\al_1^{(2)*}}{\tau_{11}l_{11}},~e^{\Del_{21}}=\frac{\vrho_{12}\al_1^{(1)}\al_1^{(2)*}\al_2^{(2)}}{\tau_{11}l_{21}},e^{\Del_{12}}=\frac{\vrho_{11}\al_1^{(1)}\al_1^{(2)}\al_1^{(2)*}}{\tau_{12}l_{12}},\nonumber\\
&&e^{\Del_{22}}=\frac{\vrho_{12}\al_1^{(1)}\al_2^{(2)}\al_2^{(2)*}}{\tau_{12}l_{22}},~e^{\Lam_{11}}=\frac{\vrho_{21}\al_2^{(1)}\al_1^{(2)}\al_1^{(2)*}}{\tau_{21}l_{11}},~e^{\Lam_{21}}=\frac{\vrho_{22}\al_2^{(1)}\al_2^{(2)}\al_1^{(2)*}}{\tau_{21}l_{21}},\nonumber\\
&&e^{\Lam_{12}}=\frac{\vrho_{21}\al_2^{(1)}\al_1^{(2)}\al_2^{(2)*}}{\tau_{22}l_{12}},~e^{\Lam_{22}}=\frac{\vrho_{22}\al_2^{(1)}\al_2^{(2)}\al_2^{(2)*}}{\tau_{22}l_{22}},e^{\Del_{13}}=\frac{\theta_1^2\al_1^{(1)}\al_1^{(1*)}\al_2^{(1)}}{\kappa_{21}\kappa_{11}},\nonumber\\
&&e^{\Del_{31}}=\frac{\theta_1^2\al_1^{(1)}\al_2^{(1)}\al_2^{(1)*}}{\kappa_{12}\kappa_{22}},~e^{\ga_{11}}=\frac{-\vrho_{11}\al_1^{(1)}\al_1^{(1*)}\al_1^{(2)}}{\tau_{11}^*\kappa_{11}},~e^{\ga_{21}}=\frac{-\vrho_{21}\al_1^{(1)*}\al_1^{(2)}\al_2^{(1)}}{\tau_{11}^*\kappa_{21}},\nonumber\\
&&e^{\ga_{12}}=\frac{-\vrho_{11}\al_1^{(1)}\al_1^{(2)}\al_2^{(1)*}}{\tau_{21}^*\kappa_{12}},~e^{\ga_{22}}=\frac{-\vrho_{21}\al_2^{(1)}\al_2^{(1)*}\al_1^{(2)}}{\tau_{21}^*\kappa_{22}},~e^{\mu_{11}}=\frac{-\vrho_{12}\al_1^{(1)}\al_1^{(1)*}\al_2^{(2)}}{\tau_{12}^*\kappa_{11}},\nonumber\\
&&e^{\mu_{21}}=\frac{-\vrho_{22}\al_1^{(1)*}\al_2^{(1)}\al_2^{(2)}}{\tau_{12}^*\kappa_{21}},~e^{\mu_{12}}=\frac{-\vrho_{12}\al_1^{(1)}\al_2^{(1)*}\al_2^{(2)}}{\tau_{22}^*\kappa_{12}},~e^{\mu_{22}}=\frac{-\vrho_{22}\al_2^{(1)}\al_2^{(1)*}\al_2^{(2)}}{\tau_{22}^*\kappa_{22}},\nonumber\\
&&e^{\ga_{13}}=\frac{\theta_2^2\al_1^{(2)}\al_1^{(2)*}\al_2^{(2)}}{l_{11}l_{21}},~e^{\ga_{31}}=\frac{\theta_2^2\al_1^{(2)}\al_2^{(2)*}\al_2^{(2)}}{l_{12}l_{22}},~e^{\del_1}=\frac{\al_1^{(1)}\al_1^{(1)*}}{\kappa_{11}},~e^{\del_2}=\frac{\al_2^{(1)}\al_1^{(1)*}}{\kappa_{21}},\nonumber\\
&&e^{\Th_{11}}=\frac{\theta_1^2\vrho_{11}\vrho_{11}^*\vrho_{21}\al_1^{(1)}\al_1^{(1)*}\al_1^{(2)}\al_1^{(2)*}\al_2^{(1)}}{\kappa_{11}\kappa_{21}\tau_{11}\tau_{11}^*\tau_{21}l_{11}},~e^{\Th_{21}}=\frac{\theta_1^2\vrho_{12}\vrho_{11}^*\vrho_{22}\al_1^{(1)}\al_1^{(1)*}\al_1^{(2)*}\al_2^{(1)}\al_2^{(2)}}{\kappa_{11}\kappa_{21}\tau_{11}\tau_{12}^*\tau_{21}l_{21}},\nonumber\\
&&e^{\Th_{12}}=\frac{\theta_1^2\vrho_{11}\vrho_{12}^*\vrho_{21}\al_1^{(1)}\al_1^{(1)*}\al_1^{(2)}\al_2^{(1)}\al_2^{(2)*}}{\kappa_{11}\kappa_{21}\tau_{12}\tau_{11}^*\tau_{22}l_{12}},~e^{\Th_{22}}=\frac{\theta_1^2\vrho_{12}\vrho_{12}^*\vrho_{22}\al_1^{(1)}\al_1^{(1)*}\al_2^{(1)}\al_2^{(2)*}\al_2^{(2)}}{\kappa_{11}\kappa_{21}\tau_{12}\tau_{12}^*\tau_{22}l_{22}},\nonumber\\
&&e^{\Phi_{11}}=\frac{\theta_1^2\vrho_{11}\vrho_{21}^*\vrho_{21}\al_1^{(1)}\al_1^{(2)*}\al_1^{(2)}\al_2^{(1)*}\al_2^{(1)}}{\kappa_{12}\kappa_{22}\tau_{11}\tau_{21}^*\tau_{21}l_{11}},~e^{\Phi_{21}}=\frac{\theta_1^2\vrho_{12}\vrho_{21}^*\vrho_{22}\al_1^{(1)}\al_1^{(2)*}\al_2^{(1)}\al_2^{(1)*}\al_2^{(2)}}{\kappa_{12}\kappa_{22}\tau_{11}\tau_{22}^*\tau_{21}l_{21}},\nonumber\\
&&e^{\Phi_{12}}=\frac{\theta_1^2\vrho_{11}\vrho_{21}\vrho_{22}^*\al_1^{(1)}\al_1^{(2)}\al_2^{(1)}\al_2^{(1)*}\al_2^{(2)*}}{\kappa_{12}\kappa_{22}\tau_{12}\tau_{21}^*\tau_{22}l_{12}},~e^{\Phi_{22}}=\frac{\theta_1^2\vrho_{12}\vrho_{22}\vrho_{22}^*\al_1^{(1)}\al_2^{(1)}\al_2^{(1)*}\al_2^{(2)}\al_2^{(2)*}}{\kappa_{12}\kappa_{22}\tau_{12}\tau_{22}^*\tau_{22}l_{22}},\nonumber\\
&&e^{\Del_{14}}=\frac{\theta_2^2\theta_2^{2*}\vrho_{11}\vrho_{12}\al_1^{(1)}\al_1^{(2)}\al_1^{(2)*}\al_2^{(2)}\al_2^{(2)*}}{\tau_{11}\tau_{12}l_{11}l_{12}l_{21}l_{22}},~e^{\Del_{41}}=\frac{\theta_2^2\theta_2^{2*}\vrho_{21}\vrho_{22}\al_1^{(2)}\al_1^{(2)*}\al_2^{(1)}\al_2^{(2)}\al_2^{(2)*}}{\tau_{21}\tau_{22}l_{11}l_{12}l_{21}l_{22}},\nonumber\\
&&e^{\nu_{11}}=\frac{-\theta_2^2\vrho_{11}\vrho_{11}^*\vrho_{12}\al_1^{(1)}\al_1^{(1)*}\al_1^{(2)}\al_1^{(2)*}\al_2^{(2)}}{\kappa_{11}l_{11}\tau_{11}\tau_{11}^*\tau_{12}^*l_{21}},~e^{\nu_{21}}=\frac{-\theta_2^2\vrho_{21}\vrho_{11}^*\vrho_{22}\al_1^{(1)*}\al_1^{(2)}\al_1^{(2)*}\al_2^{(1)}\al_2^{(2)}}{\kappa_{21}\tau_{21}\tau_{11}^*\tau_{12}^*l_{11}l_{21}},
\end{eqnarray*}
\begin{eqnarray*}
&&e^{\nu_{12}}=\frac{-\theta_2^2\vrho_{11}\vrho_{21}^*\vrho_{12}\al_1^{(1)}\al_1^{(2)*}\al_1^{(2)}\al_2^{(1)*}\al_2^{(2)}}{\kappa_{12}\tau_{11}\tau_{21}^*\tau_{22}^*l_{11}l_{21}},~e^{\nu_{22}}=\frac{-\theta_2^2\vrho_{21}\vrho_{21}^*\vrho_{22}\al_1^{(2)}\al_1^{(2)*}\al_2^{(1)}\al_2^{(1)*}\al_2^{(2)}}{\kappa_{22}\tau_{21}\tau_{21}^*\tau_{22}^*l_{11}l_{21}},\nonumber\\
&&e^{\chi_{11}}=\frac{-\theta_2^2\vrho_{11}\vrho_{12}^*\vrho_{12}\al_1^{(1)}\al_1^{(1)*}\al_1^{(2)}\al_2^{(2)*}\al_2^{(2)}}{\kappa_{11}\tau_{11}^*\tau_{12}^*\tau_{12}l_{12}l_{22}},~e^{\chi_{21}}=\frac{-\theta_2^2\vrho_{21}\vrho_{12}^*\vrho_{22}\al_2^{(1)}\al_1^{(1)*}\al_1^{(2)}\al_2^{(2)*}\al_2^{(2)}}{\kappa_{21}\tau_{11}^*\tau_{12}^*\tau_{22}l_{12}l_{22}},\\
&&e^{\chi_{12}}=\frac{-\theta_2^2\vrho_{11}\vrho_{22}^*\vrho_{12}\al_1^{(1)}\al_2^{(1)*}\al_1^{(2)}\al_2^{(2)*}\al_2^{(2)}}{\kappa_{12}\tau_{21}^*\tau_{22}^*\tau_{12}l_{12}l_{22}},~e^{\chi_{22}}=\frac{-\theta_2^2\vrho_{21}\vrho_{22}^*\vrho_{22}\al_2^{(1)}\al_2^{(1)*}\al_1^{(2)}\al_2^{(2)*}\al_2^{(2)}}{\kappa_{22}\tau_{21}^*\tau_{22}^*\tau_{22}l_{12}l_{22}},\\
&&e^{\ga_{14}}=\frac{\theta_1^2\theta_1^{2*}\vrho_{11}\vrho_{21}\al_1^{(1)}\al_1^{(2)}\al_1^{(1)*}\al_2^{(1)}\al_2^{(1)*}}{\tau_{11}^*\tau_{12}^*\kappa_{11}\kappa_{12}\kappa_{21}\kappa_{22}},~e^{\ga_{41}}=\frac{\theta_1^2\theta_1^{2*}\vrho_{12}\vrho_{22}\al_1^{(1)}\al_2^{(1)}\al_1^{(1)*}\al_2^{(1)*}\al_2^{(2)}}{\tau_{12}^*\tau_{22}^*\kappa_{11}\kappa_{12}\kappa_{21}\kappa_{22}},\\
&&e^{\Del_{15}}=\frac{\theta_1^2\theta_2^2\theta_2^{2*}\vrho_{11}\vrho_{21}\vrho_{12}\vrho_{22}\vrho_{11}^*\vrho_{12}^*\al_1^{(1)}\al_1^{(1)*}\al_1^{(2)}\al_1^{(2)*}\al_2^{(1)}\al_2^{(2)}\al_2^{(2)*}}{\kappa_{11}\kappa_{21}\tau_{11}\tau_{11}^*\tau_{21}\tau_{12}\tau_{12}^*\tau_{22}l_{11}l_{12}l_{21}l_{22}},~e^{\del_3}=\frac{|\al_2^{(1)}|^2}{\kappa_{22}},\\
&&e^{\Del_{51}}=\frac{\theta_1^2\theta_2^2\theta_2^{2*}\vrho_{11}\vrho_{21}\vrho_{12}\vrho_{22}\vrho_{21}^*\vrho_{22}^*\al_1^{(1)}\al_1^{(2)*}\al_1^{(2)}\al_2^{(1)*}\al_2^{(1)}\al_2^{(2)}\al_2^{(2)*}}{\kappa_{12}\kappa_{22}\tau_{11}\tau_{21}^*\tau_{21}\tau_{12}\tau_{22}^*\tau_{22}l_{11}l_{12}l_{21}l_{22}},~e^{\psi_{11}}=\frac{|\al_1^{(2)}|^2}{l_{11}},\\
&&e^{\ga_{15}}=\frac{\theta_1^2\theta_2^2\theta_1^{2*}\vrho_{11}\vrho_{21}\vrho_{12}\vrho_{22}\vrho_{21}^*\vrho_{12}^*\al_1^{(1)}\al_1^{(1)*}\al_1^{(2)}\al_1^{(2)*}\al_2^{(1)}\al_2^{(2)}\al_2^{(1)*}}{\kappa_{11}\kappa_{21}\kappa_{12}\kappa_{22}\tau_{11}\tau_{11}^*\tau_{21}\tau_{12}^*\tau_{22}^*\tau_{21}^*l_{11}l_{21}},~e^{\psi_{21}}=\frac{\al_1^{(2)*}\al_2^{(2)}}{l_{21}},\\
&&e^{\ga_{51}}=\frac{\theta_1^2\theta_2^2\theta_1^{2*}\vrho_{11}\vrho_{21}\vrho_{12}\vrho_{22}\vrho_{22}^*\vrho_{12}^*\al_1^{(1)}\al_1^{(1)*}\al_1^{(2)}\al_2^{(1)*}\al_2^{(1)}\al_2^{(2)}\al_2^{(2)*}}{\kappa_{11}\kappa_{21}\kappa_{12}\kappa_{22}\tau_{22}\tau_{11}^*\tau_{12}\tau_{12}^*\tau_{22}^*\tau_{21}^*l_{12}l_{22}},~e^{\psi_{12}}=\frac{\al_1^{(2)}\al_2^{(2)*}}{l_{12}},\\
&&e^{\del_{18}}=\frac{\theta_1^2\theta_2^2\theta_1^{2*}\theta_2^{2*}\vrho_{11}\vrho_{21}\vrho_{12}\vrho_{22}\vrho_{11}^*\vrho_{12}^*\vrho_{21}^*\vrho_{22}^*|\al_1^{(1)}|^2|\al_1^{(2)}|^2|\al_2^{(1)}|^2|\al_2^{(2)}|^2}{\kappa_{11}\kappa_{21}\kappa_{12}\kappa_{22}|\tau_{11}|^2|\tau_{12}|^2|\tau_{21}|^2 |\tau_{22}|^2l_{11}l_{21}l_{12}l_{22}},~e^{\psi_{22}}=\frac{|\al_2^{(2)}|^2}{l_{22}},\\
&&e^{\del_{11}}=\frac{|\vrho_{11}|^2|\al_1^{(1)}|^2|\al_1^{(2)}|^2}{\kappa_{11}|\tau_{11}|^2l_{11}},~e^{\del_{21}}=\frac{\vrho_{12}\vrho_{11}^*\al_1^{(1)}\al_1^{(1)*}\al_2^{(2)}\al_1^{(2)*}}{\kappa_{11}\tau_{11}\tau_{12}^*l_{21}},~e^{\del_{12}}=\frac{\vrho_{11}\vrho_{12}^*\al_1^{(1)}\al_1^{(1)*}\al_1^{(2)}\al_2^{(2)*}}{\kappa_{11}\tau_{11}^*\tau_{12}l_{12}},\\
&&e^{\del_{22}}=\frac{|\vrho_{12}|^2|\al_1^{(1)}|^2|\al_2^{(2)}|^2}{\kappa_{11}|\tau_{12}|^2l_{22}},~e^{\vth_{11}}=\frac{\vrho_{21}\vrho_{11}^*\al_1^{(1)*} |\al_1^{(2)}|^2 \al_2^{(1)}}{\kappa_{21}\tau_{11}^*\tau_{21}l_{11}},~e^{\vth_{21}}=\frac{\vrho_{22}\vrho_{11}^*\al_1^{(1)*}\al_1^{(2)*}\al_2^{(1)} \al_2^{(2)}}{\kappa_{21}\tau_{12}^*\tau_{21}l_{21}},\\
&&e^{\vth_{12}}=\frac{\vrho_{21}\vrho_{12}^*\al_1^{(1)*}\al_1^{(2)}\al_2^{(1)} \al_2^{(2)*}}{\kappa_{21}\tau_{11}^*\tau_{22}l_{12}},~e^{\vth_{22}}=\frac{\vrho_{22}\vrho_{12}^*\al_1^{(1)*}\al_2^{(2)}\al_2^{(1)} \al_2^{(2)*}}{\kappa_{21}\tau_{12}^*\tau_{22}l_{22}},~e^{\vphi_{11}}=\frac{\vrho_{11}\vrho_{21}^*\al_1^{(1)}\al_1^{(2)}\al_1^{(2)*} \al_2^{(1)*}}{\kappa_{12}\tau_{21}^*\tau_{11}l_{11}},\\
&&e^{\vphi_{21}}=\frac{\vrho_{12}\vrho_{21}^*\al_1^{(1)}\al_1^{(2)*}\al_2^{(1)*} \al_2^{(2)}}{\kappa_{12}\tau_{22}^*\tau_{11}l_{21}},~e^{\vphi_{12}}=\frac{\vrho_{11}\vrho_{22}^*\al_1^{(1)}\al_1^{(2)}\al_2^{(2)*} \al_2^{(1)*}}{\kappa_{12}\tau_{21}^*\tau_{12}l_{12}},~e^{\vphi_{22}}=\frac{\vrho_{12}\vrho_{22}^*\al_1^{(1)}\al_2^{(2)}\al_2^{(2)*} \al_2^{(1)*}}{\kappa_{12}\tau_{22}^*\tau_{12}l_{22}},\\
&&e^{\vsa_{11}}=\frac{\vrho_{21}\vrho_{21}^*\al_1^{(2)}\al_1^{(2)*}\al_2^{(1)} \al_2^{(1)*}}{\kappa_{12}\tau_{21}^*\tau_{21}l_{11}},~e^{\vsa_{21}}=\frac{\vrho_{22}\vrho_{21}^*\al_1^{(2)*}\al_2^{(1)*}\al_2^{(1)} \al_2^{(2)}}{\kappa_{22}\tau_{22}^*\tau_{21}l_{21}},~e^{\vsa_{12}}=\frac{\vrho_{21}\vrho_{22}^*\al_1^{(2)}\al_2^{(2)*}\al_2^{(1)} \al_2^{(1)*}}{\kappa_{22}\tau_{21}^*\tau_{22}l_{12}},\\
&&e^{\vsa_{22}}=\frac{\vrho_{22}\vrho_{22}^*\al_2^{(2)}\al_2^{(2)*}\al_2^{(1)} \al_2^{(1)*}}{\kappa_{22}\tau_{22}^*\tau_{22}l_{22}},~e^{\del_{17}}=\frac{\theta_1^2\theta_1^{2*}|\al_1^{(1)}|^2|\al_2^{(1)}|^2}{\kappa_{12}\kappa_{21}\kappa_{11}\kappa_{22}},~e^{\del_{19}}=\frac{\theta_2^2\theta_2^{2*}|\al_1^{(2)}|^2|\al_2^{(2)}|^2}{l_{12}l_{21}l_{11}l_{22}},\\
&&e^{\phi_{11}}=\frac{\theta_1^2\theta_1^{2*}\vrho_{11}\vrho_{21}\vrho_{11}^*\vrho_{21}^*|\al_1^{(1)}|^2|\al_1^{(2)}|^2|\al_2^{(1)}|^2}{\kappa_{12}\kappa_{21}\kappa_{11}\kappa_{22}|\tau_{11}|^2|\tau_{21}|^2l_{11}},~e^{\phi_{21}}=\frac{\theta_1^2\theta_1^{2*}\vrho_{12}\vrho_{22}\vrho_{11}^*\vrho_{21}^*|\al_1^{(1)}|^2\al_1^{(2)*}|\al_2^{(1)}|^2\al_2^{(2)}}{\kappa_{12}\kappa_{21}\kappa_{11}\kappa_{22}\tau_{11} \tau_{21} \tau_{12}^* \tau_{22}^* l_{21}},\\
&&e^{\phi_{12}}=\frac{\theta_1^2\theta_1^{2*}\vrho_{11}\vrho_{21}\vrho_{12}^*\vrho_{22}^*|\al_1^{(1)}|^2\al_1^{(2)}|\al_2^{(1)}|^2\al_2^{(2)*}}{\kappa_{12}\kappa_{21}\kappa_{11}\kappa_{22}\tau_{11}^* \tau_{21}^* \tau_{12} \tau_{22} l_{12}},~e^{\phi_{22}}=\frac{\theta_1^2\theta_1^{2*}\vrho_{12}\vrho_{22}\vrho_{12}^*\vrho_{22}^*|\al_1^{(1)}|^2|\al_2^{(2)}|^2|\al_2^{(1)}|^2}{\kappa_{12}\kappa_{21}\kappa_{11}\kappa_{22}|\tau_{12}|^2|\tau_{22}|^2l_{22}},\\
&&e^{\del_{13}}=\frac{\theta_2^2\theta_2^{2*}\vrho_{11}\vrho_{12}\vrho_{11}^*\vrho_{12}^*|\al_1^{(1)}|^2|\al_1^{(2)}|^2|\al_2^{(2)}|^2}{\kappa_{11}l_{11}l_{21}l_{12}l_{22}|\tau_{11}|^2|\tau_{12}|^2},~e^{\del_{14}}=\frac{\theta_2^2\theta_2^{2*}\vrho_{21}\vrho_{22}\vrho_{11}^*\vrho_{12}^* \al_1^{(1)*} |\al_1^{(2)}|^2 \al_2^{(1)} |\al_2^{(2)}|^2}{\kappa_{21}l_{11}l_{21}l_{12}l_{22}\tau_{11}^*\tau_{12}^*\tau_{21}\tau_{22}},\\
&&e^{\del_{15}}=\frac{\theta_2^2\theta_2^{2*}\vrho_{11}\vrho_{12}\vrho_{21}^*\vrho_{22}^* \al_1^{(1)} |\al_1^{(2)}|^2 \al_2^{(1)*} |\al_2^{(2)}|^2}{\kappa_{12}l_{11}l_{21}l_{12}l_{22}\tau_{21}^*\tau_{22}^*\tau_{11}\tau_{12}},~e^{\del_{16}}=\frac{\theta_2^2\theta_2^{2*}\vrho_{21}\vrho_{22}\vrho_{21}^*\vrho_{22}^* \al_1^{(2)*} |\al_2^{(1)}|^2 |\al_2^{(2)}|^2}{\kappa_{22}l_{11}l_{21}l_{12}l_{22}\tau_{21}^*\tau_{22}^*\tau_{21}\tau_{22}},
\end{eqnarray*}
\begin{eqnarray*}
&&\theta_{1}=(k_1-k_2),~\theta_{2}=(l_1-l_2), ~\vrho_{nm}=(k_n-l_m), ~\tau_{nm}=(k_n+l_m^*),\\
&&l_{nm}=(l_n+l_m^*)^2,~ \kappa_{nm}=(k_n+k_m^*)^2, ~n,m=1,2.
\end{eqnarray*}
\section{S.2. Degenerate two-soliton solution}
In the limit, $k_{1}=l_{1}$ and $k_{2}=l_{2}$, the above given nondegenerate two-soliton solution  is reduced to the following standard degenerate two-soliton solution \cite{j}, that is 
\begin{equation}
q_{j}(x,t)= \frac{\alpha_{1}^{(j)}e^{\eta_{1}}+\alpha_{2}^{(j)}e^{\eta_{2}}+e^{\eta_{1}+\eta_{1}^*+\eta_{2}+\delta_{1j}}+e^{\eta_{1}+\eta_{2}+\eta_{2}^*+\delta_{2j}}}{1+e^{\eta_{1}+\eta_{1}^*+R_{1}}+e^{\eta_{1}+\eta_{2}^*+\delta_{0}}+e^{\eta_{1}^*+\eta_{2}+\delta_{0}^*}+e^{\eta_{2}+\eta_{2}^*+R_{2}}+e^{\eta_{1}+\eta_{1}^*+\eta_{2}+\eta_{2}^*+R_{3}}},
\end{equation}
where 

$j=1,2$, $\eta_{j}=k_{j}(t+ik_{jz})$, $e^{\delta_{0}}=\frac{k_{12}}{k_{1}+k_{2}^*}$, $e^{R_{1}}=\frac{k_{11}}{k_{1}+k_{1}^*}$, $e^{R_{2}}=\frac{k_{22}}{k_{2}+k_{2}^*}$,\\
$e^{\delta_{1j}}=\frac{(k_{1}-k_{2})(\alpha_{1}^{(j)}k_{21}-\alpha_{2}^{(j)}k_{11})}{(k_{1}+k_{1}^*)(k_{1}^*+k{2})}$, $e^{\delta_{2j}}=\frac{(k_{2}-k_{1})(\alpha_{2}^{(j)}k_{12}-\alpha_{1}^{(j)}k_{22})}{(k_{2}+k_{2}^*)(k_{1}+k{2}^*)}$,\\
$e^{R_{3}}=\frac{|k_{1}-k_{2}|^2}{(k_{1}+k_{1}^*)(k_{2}+k_{2}^*)|k_{1}+k_{2}^*|^2}(k_{11}k_{22}-k_{12}k_{21})$ and
$k_{il}=\frac{\mu~\sum_{n=1}^{2}~\alpha_{i}^{(n)}\alpha_{i}^{(n)^*}}{(k_{i}+k_{l}^*)}$; $i,l=1,2$.

The energy sharing collision of the above degenerate two-soliton solution was studied in \cite{j}.
\section{S.3. Asymptotic analysis: Shape preserving collision of double hump soliton}
To study the interaction between nondegenerate solitons, we carefully perform the asymptotic analysis for the nondegenerate two soliton solution (2a)-(2c). Here, we present the asymptotic analysis for the shape preserving collision that occur between the double hump solitons. To explore this, we consider the choice of wave parameters as $k_{1R}>l_{1R}$, $k_{2R}<l_{2R}$, $k_{1I}>k_{2I}$, $l_{1I}>l_{2I}$, $k_{1R}>k_{2R}$, $l_{1R}<l_{2R}$, $k_{1I}=l_{1I}$ and $k_{2I}=l_{2I}$. Under this parametric choice, we deduce the following asymptotic forms corresponding to two individual double hump solitons in both the modes before and after collision can be given as follows.\\ 
\underline{\textit{Before collision}}: $z\rightarrow -\infty$\\
\underline{soliton 1}: In this limit, the asymptotic forms deduced from the two soliton solution for soliton 1 in both the modes before collision,
\begin{subequations}
\begin{eqnarray}
&&q_{1}=\frac{2A_1^{1-}k_{1R}e^{i\eta_{1I}}\cosh(\xi_{1R}+\frac{\phi_{1}}{2})}{\big[{\frac{(k_{1}^{*}-l_{1}^{*})^{\frac{1}{2}}}{(k_{1}^{*}+l_{1})^{\frac{1}{2}}}}\cosh(\eta_{1R}+\xi_{1R}+\frac{\delta_{1}^{(1)}}{2})+\frac{(k_{1}+l_{1}^{*})^{\frac{1}{2}}}{(k_{1}-l_{1})^{\frac{1}{2}}}\cosh(\eta_{1R}-\xi_{1R}+\frac{\delta_{1}-\delta_{4}}{2})\big]},\\
&&q_{2}=\frac{2A_1^{2-}l_{1R}e^{i\xi_{1I}}\cosh(\eta_{1R}++\frac{\phi_{2}}{2})}{\big[\frac{i(k_{1}^{*}-l_{1}^{*})^{\frac{1}{2}}}{(k_{1}+l_{1}^{*})^{\frac{1}{2}}}\cosh(\eta_{1R}+\xi_{1R}+\frac{\delta_{1}^{(1)}}{2})+\frac{(k_{1}^{*}+l_{1})^{1/2}}{(l_{1}-k_{1})^{1/2}}\cosh(\eta_{1R}-\xi_{1R}+\frac{\delta_{1}-\delta_{4}}{2})\big]}.
\end{eqnarray}\end{subequations}\\
\underline{soliton 2}: The asymptotic expressions for the soliton 2 in the two modes before collision turn out to be,
\begin{subequations}
\begin{eqnarray}
&&q_{1}=\frac{2k_{2R}A_2^{1-}(k_{1}-k_{2})(k_{2}-l_{1})^{\frac{1}{2}}(k_{1}+k_{2}^{*})(k_{2}^{*}+l_{1})^{\frac{1}{2}}e^{i\eta_{2I}}\cosh(\xi_{2R}+\frac{\varphi_{1}}{2})}{(k_{1}^{*}-k_{2}^{*})(k_{2}^{*}-l_{1}^{*})^{\frac{1}{2}}(k_{1}^{*}+k_{2})(k_{2}+l_{1}^{*})^{\frac{1}{2}}\Gamma_1},\\
&&q_2=\frac{2l_{2R}A_2^{2-}(l_{1}-l_{2})(k_{1}-l_{2})^{\frac{1}{2}}(k_{1}+l_{2}^{*})^{\frac{1}{2}}(l_{1}+l_{2}^{*})e^{i\xi_{2I}}\cosh(\eta_{2R}+\frac{\varphi_{2}}{2})}{(k_{1}^{*}-l_{2}^{*})^{\frac{1}{2}}(l_{1}^{*}-l_{2}^{*})(k_{1}^{*}+l_{2})^{\frac{1}{2}}(l_{1}^{*}+l_{2})\Gamma_2}.
\end{eqnarray} \end{subequations}
In the above,
\begin{eqnarray*}
\Gamma_1=\big[\frac{(k_{2}^{*}-l_{2}^{*})^{\frac{1}{2}}}{(k_{2}^{*}+l_{2})^{\frac{1}{2}}}\cosh(\eta_{2R}+\xi_{2R}+\frac{\gamma_{1}^{(1)}-\delta_{1}^{(1)}}{2})+\frac{(k_{2}+l_{2}^{*})^{\frac{1}{2}}}{(k_{2}-l_{2})^{\frac{1}{2}}}\cosh(\eta_{2R}-\xi_{2R}+\frac{\gamma_1-\gamma_5}{2})\big],\\
\Gamma_2=\big[\frac{(k_{2}^{*}-l_{2}^{*})^{\frac{1}{2}}}{(k_{1}+l_{2}^*)^{\frac{1}{2}}}\cosh(\eta_{2R}+\xi_{2R}+\frac{\gamma_{1}^{(1)}-\delta_{1}^{(1)}}{2})+\frac{(k_{2}^*+l_{2})^{\frac{1}{2}}}{(k_{2}-l_{2})^{\frac{1}{2}}}\cosh(\eta_{2R}-\xi_{2R}+\frac{\gamma_1-\gamma_5}{2})\big].
\end{eqnarray*}
\\
\underline{\textit{After collision}}: $z\rightarrow +\infty$\\
\underline{soliton 1}: The asymptotic forms for soliton 1 after collision deduced as,
\begin{subequations}
\begin{eqnarray}
&&q_{1}=\frac{2k_{1R}A_1^{1+}(k_{1}-k_{2})(k_{1}-l_{2})^{\frac{1}{2}}(k_{1}^*+k_{2})(k_{1}^{*}+l_{2})^{\frac{1}{2}}e^{i\eta_{1I}}\cosh(\xi_{1R}+\frac{\hat{\phi}_{1}}{2})}{(k_{1}^{*}-k_{2}^{*})(k_{1}^{*}-l_{2}^{*})^{\frac{1}{2}}(k_{1}+k_{2}^*)(k_{1}+l_{2}^{*})^{\frac{1}{2}}\Gamma_3},\\
&&q_2=\frac{2l_{1R}A_1^{2+}(l_{1}-l_{2})(k_{2}-l_{1})^{\frac{1}{2}}(k_{2}+l_{1}^{*})^{\frac{1}{2}}(l_{1}^*+l_{2})e^{i\xi_{1I}}\cosh(\eta_{1R}+\frac{\hat{\phi}_{2}}{2})}{(k_{2}^{*}-l_{1}^{*})^{\frac{1}{2}}(l_{1}^{*}-l_{2}^{*})(k_{2}^{*}+l_{1})^{\frac{1}{2}}(l_{1}+l_{2}^*)\Gamma_4}.
\end{eqnarray} \end{subequations}
Here, 
\begin{eqnarray*}
\Gamma_3=\big[\frac{(k_{1}^{*}-l_{1}^{*})^{\frac{1}{2}}}{(k_{1}^{*}+l_{1})^{\frac{1}{2}}}\cosh(\eta_{1R}+\xi_{1R}+\frac{\gamma_{1}^{(1)}-\delta_{2}^{(1)}}{2})+\frac{(k_{1}+l_{1}^{*})^{\frac{1}{2}}}{(k_{1}-l_{1})^{\frac{1}{2}}}\cosh(\eta_{1R}-\xi_{1R}+\frac{\gamma_4-\gamma_5}{2})\big],\\
\Gamma_4=\big[\frac{(k_{1}^{*}-l_{1}^{*})^{\frac{1}{2}}}{(k_{1}+l_{1}^*)^{\frac{1}{2}}}\cosh(\eta_{1R}+\xi_{1R}+\frac{\gamma_{1}^{(1)}-\delta_{2}^{(1)}}{2})+\frac{(k_{1}^*+l_{1})^{\frac{1}{2}}}{(k_{1}-l_{1})^{\frac{1}{2}}}\cosh(\eta_{1R}-\xi_{1R}+\frac{\gamma_4-\gamma_5}{2})\big].
\end{eqnarray*}
\underline{soliton 2}: The expression for soliton 2 after collision deduced from the two soliton solution is,
\begin{subequations}
\begin{eqnarray}
&&q_{1}=\frac{2A_2^{1+}k_{2R}e^{i\eta_{2I}}\cosh(\xi_{2R}+\frac{\hat{\varphi}_{1}}{2})}{\big[{\frac{(k_{2}^{*}-l_{2}^{*})^{\frac{1}{2}}}{(k_{2}^{*}+l_{2})^{\frac{1}{2}}}}\cosh(\eta_{2R}+\xi_{2R}+\frac{\delta_{2}^{(6)}}{2})+\frac{(k_{2}+l_{2}^{*})^{\frac{1}{2}}}{(k_{2}-l_{2})^{\frac{1}{2}}}\cosh(\eta_{2R}-\xi_{2R}+\frac{\delta_{3}-\delta_{6}}{2})\big]},\\
&&q_{2}=\frac{2A_2^{2+}l_{2R}e^{i\xi_{2I}}\cosh(\eta_{2R}++\frac{\hat{\varphi}_{2}}{2})}{\big[\frac{i(k_{2}^{*}-l_{2}^{*})^{\frac{1}{2}}}{(k_{2}+l_{2}^{*})^{\frac{1}{2}}}\cosh(\eta_{2R}+\xi_{2R}+\frac{\delta_{2}^{(6)}}{2})+\frac{(k_{2}^{*}+l_{2})^{\frac{1}{2}}}{(l_{2}-k_{2})^{\frac{1}{2}}}\cosh(\eta_{2R}-\xi_{2R}+\frac{\delta_{1}-\delta_{4}}{2})\big]},
\end{eqnarray} \end{subequations}
where $\eta_{jR}=k_{jR}(t-2k_{jI}z)$, $\eta_{jI}=k_{jI}t+(k_{jR}^{2}-k_{jI}^{2})z$, $\xi_{jR}=l_{jR}(t-2l_{jI}z)$, $\xi_{jI}=l_{jI}t+(l_{jR}^{2}-l_{jI}^{2})z$, $j=1,2$, and other constants appearing in the above asymptotic expressions are given in two soliton solution.
\begin{figure}
\centering
\includegraphics[width=0.35\linewidth]{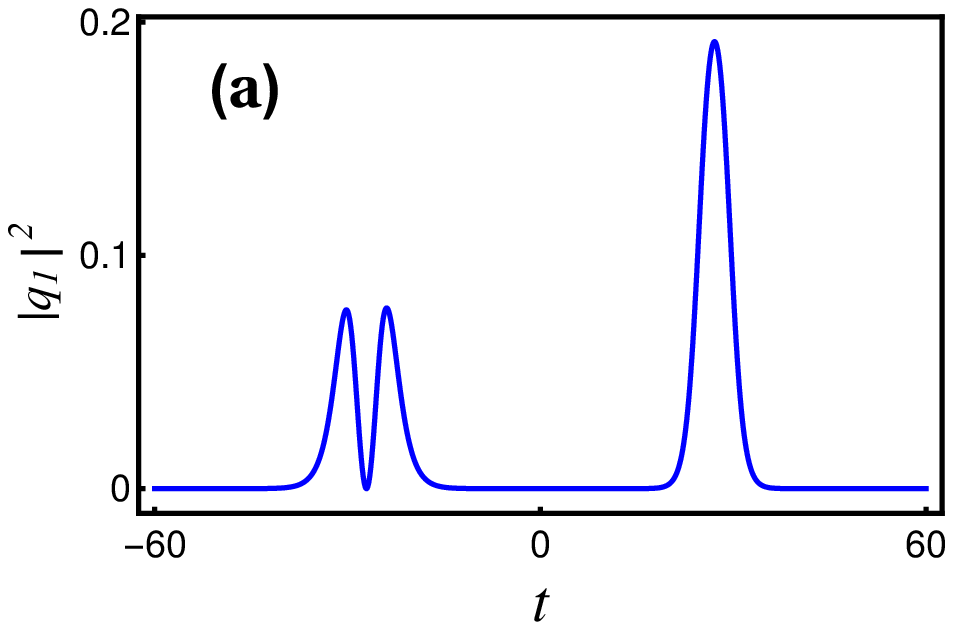}~\includegraphics[width=0.35\linewidth]{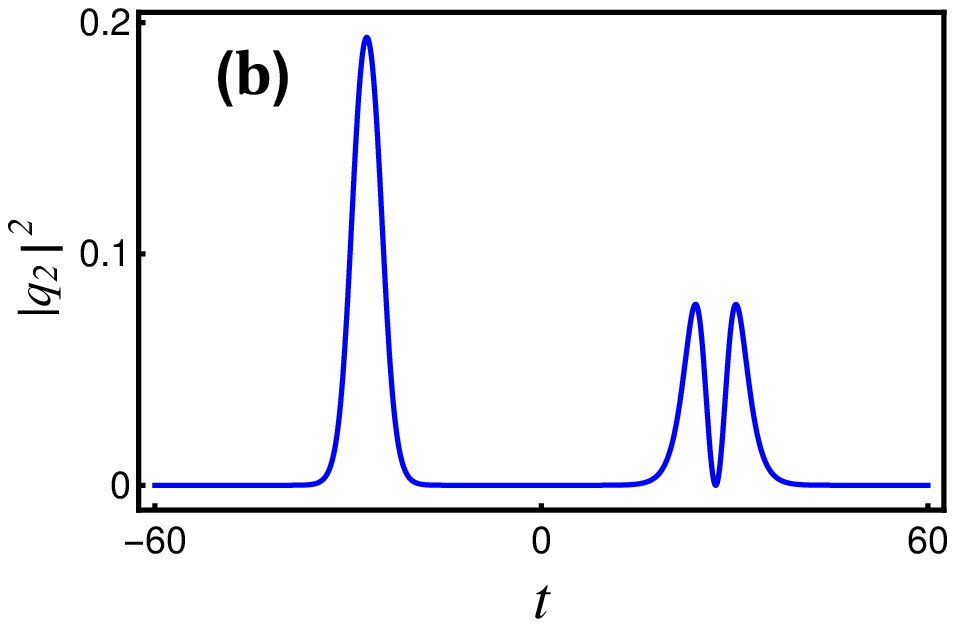}
\includegraphics[width=0.35\linewidth]{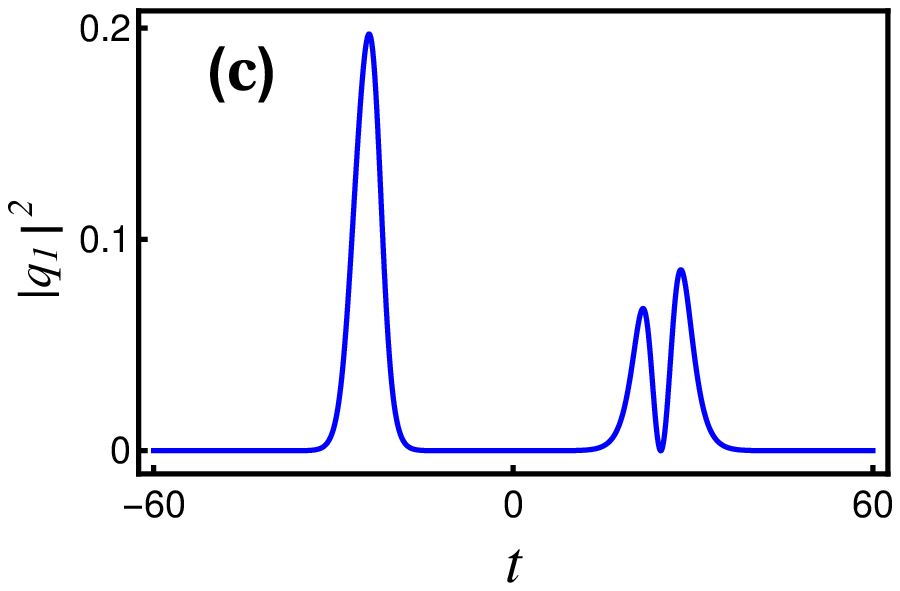}~\includegraphics[width=0.35\linewidth]{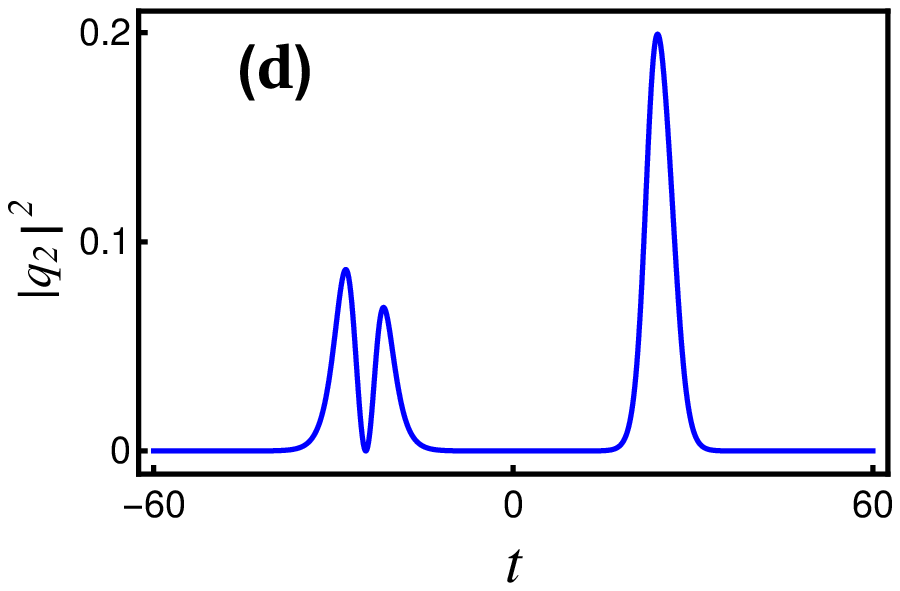}
\caption{Stability of double hump solitons-alteration of shape of the symmetric double hump soliton in both the modes: (a) and (b) denote the symmetric double hump soliton present in both the modes before collision. (c) and (d) exhibit that a slight asymmetry occur in the double hump soliton profiles (though the individual energies in each modes remains constant) after collision with symmetric single hump solitons. Note that the single hump soliton retains it shape. The parameter values are $k_{1} =0.55+0.5i$, $l_1=0.333+0.5i$, $k_{2} =-0.33+0.5i$, $l_2=-0.55-0.5i$, $\alpha_{1}^{(1)} =0.5 +0.5 i$, $\alpha_{2}^{(1)} =-0.45+0.5i$, $\alpha_{1}^{(2)}=0.45+0.5i$ and $\alpha_{2}^{(2)}=-0.5+0.5i$.}
\label{f3}
\end{figure} 
 \section{S.4. Alteration in the shape of the symmetric double hump soliton while interacting with a symmetric single hump soliton (without change in energies)}
In the above, we have analysed the the shape preserving collision that occurs between the two double hump solitons \cite{k}. However, we also come across a slightly different collision scenario while testing the stability property of a symmetric double hump soliton interacting with a symmetric single hump soliton through analytical and numerical analysis. To analyse this, we allow the double hump soliton to interact with a symmetric single hump soliton. As we have demonstrated in Fig. 1 in this supplement, by choosing appropriate initial conditions from our explicit analytical two-soliton solution, we locate these two solitons initially at well defined  separation distance. Both the symmetric double hump and single  hump solitons propagate steadily until they get disturbed by each other. During this interaction process, the double hump soliton in both the modes alone experiences a strong perturbation due to the collision with the symmetric single hump soliton. The result of their collision is reflected only in changing the shape of the symmetric double hump soliton slightly into an asymmetric form. However, the energies of both the double hump and single hump solitons do not change after collision. We also confirm this collision scenario by carrying out a detailed asymptotic analysis as has been done for the double hump soliton earlier and calculating the transition amplitudes and conservation of energy. We will present a detailed analysis of the  above dynamics in the extended version. We also note that if one sets a slight asymmetry in the intensity profile of the double hump soliton in both the modes for suitable choice of initial conditions, the asymmetric double hump becomes symmetric when it collides with the symmetric single hump soliton, again without change in energy.
\section{ S.5. Symmetric soliton profiles of nondegenerate one soliton}
\begin{figure}
\centering
\includegraphics[width=0.35\linewidth]{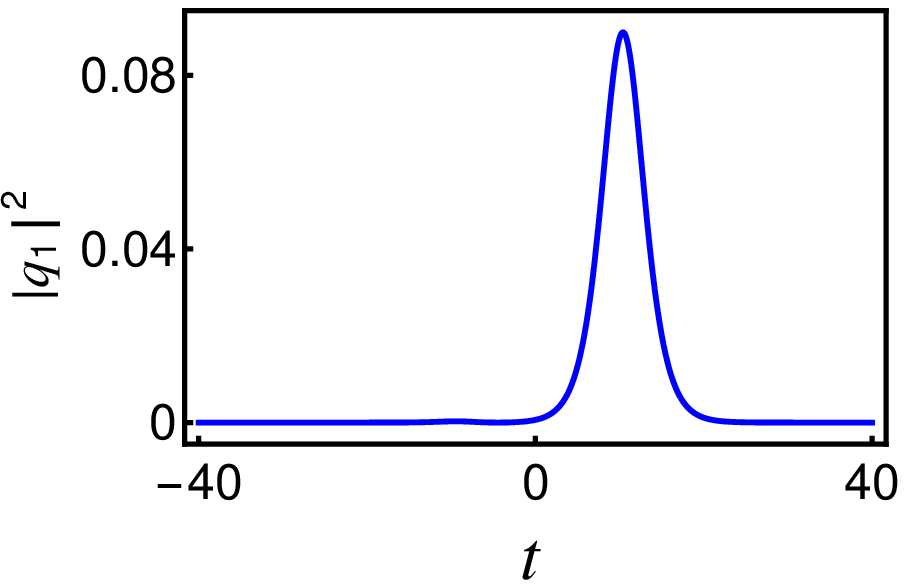}
\includegraphics[width=0.35\linewidth]{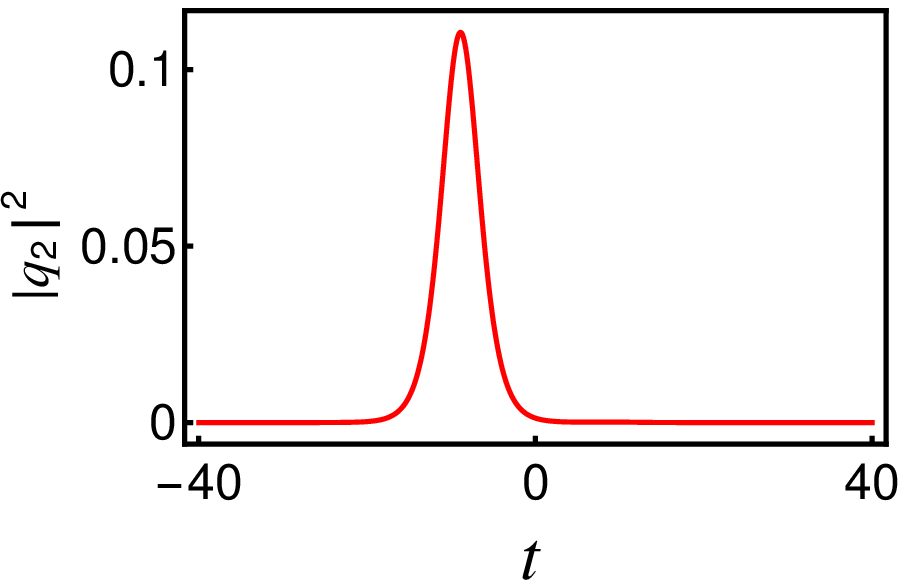}
\caption{Symmetric single hump-single hump soliton: Parameter values are $k_{1} = -0.3+ 0.5i$, $l_{1} = 0.333 + 0.5i$, $\alpha_{1}^{(1)} = 0.5 + 0.5i$ and $\alpha_{1}^{(2)} = 0.45 + 0.5i$. }
\label{f2}
\end{figure}

 We display all the four symmetric wave profiles of nondegenerate one soliton of Manakov system in Figs. (\ref{f2})-(\ref{f5}) for appropriate choice of parameters as given in the corresponding figure captions.
\begin{figure}
\centering
\includegraphics[width=0.35\linewidth]{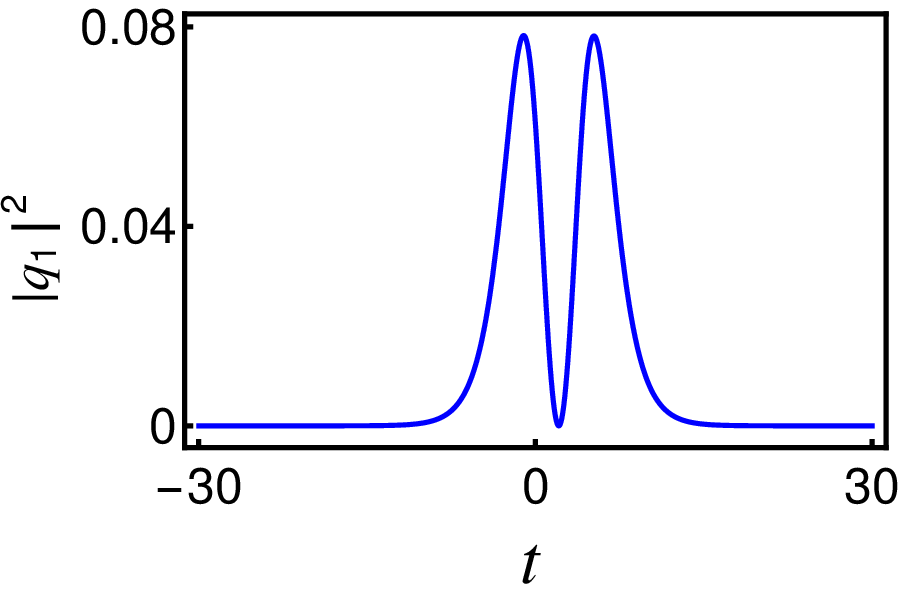}
\includegraphics[width=0.35\linewidth]{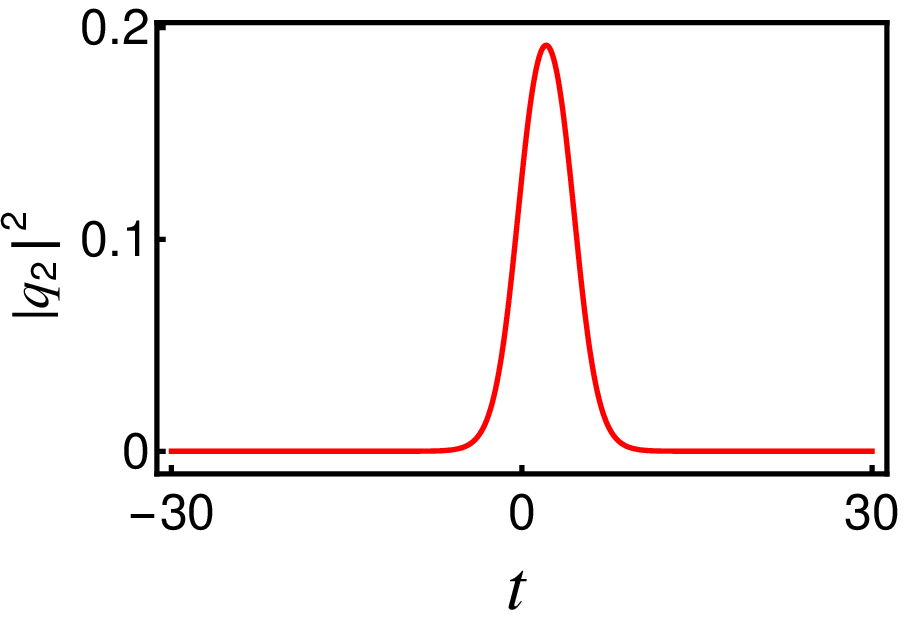}
\caption{Symmetric double hump-single hump soliton:  Parameter values are $k_{1} = 0.333+ 0.5i$, $l_{1} = 0.55 + 0.5i$, $\alpha_{1}^{(1)} = 0.5 + 0.45i$ and $\alpha_{1}^{(2)} = 0.5 + 0.5i$.}
\label{f3}
\end{figure}
\begin{figure}
\centering
\includegraphics[width=0.35\linewidth]{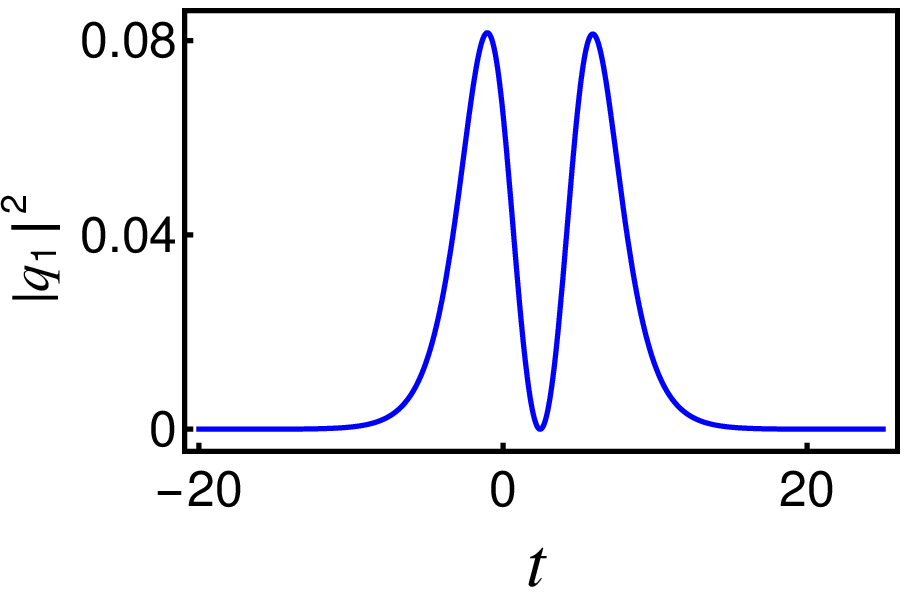}
\includegraphics[width=0.35\linewidth]{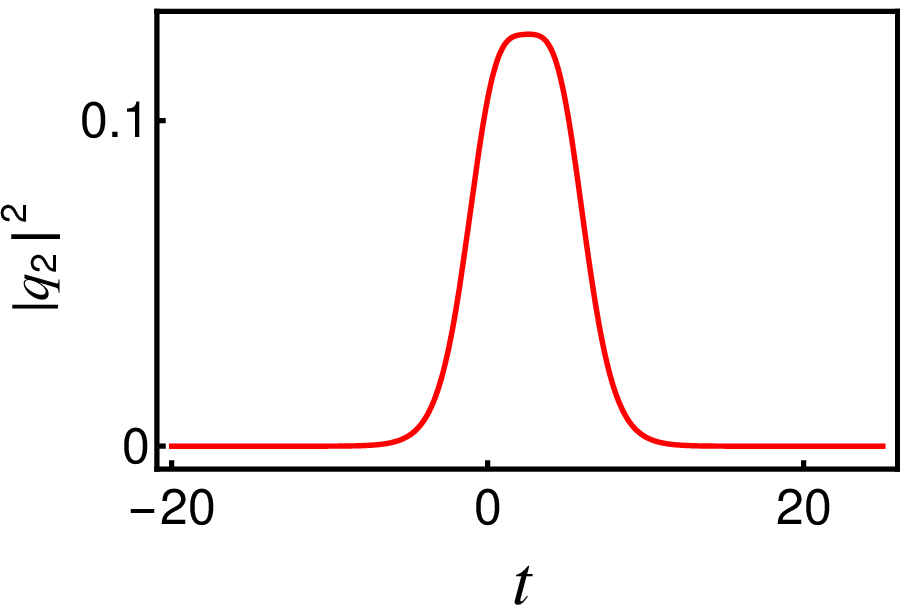}
\caption{Symmetric double hump-flat top soliton: Parameter values are $k_{1} = 0.35+ 0.5i$, $l_{1} = 0.499 + 0.5i$, $\alpha_{1}^{(1)} = 0.5 + 0.51i$ and $\alpha_{1}^{(2)} = 0.5 + 0.5i$.}
\label{f4}
\end{figure}

\begin{figure}
\centering
\includegraphics[width=0.35\linewidth]{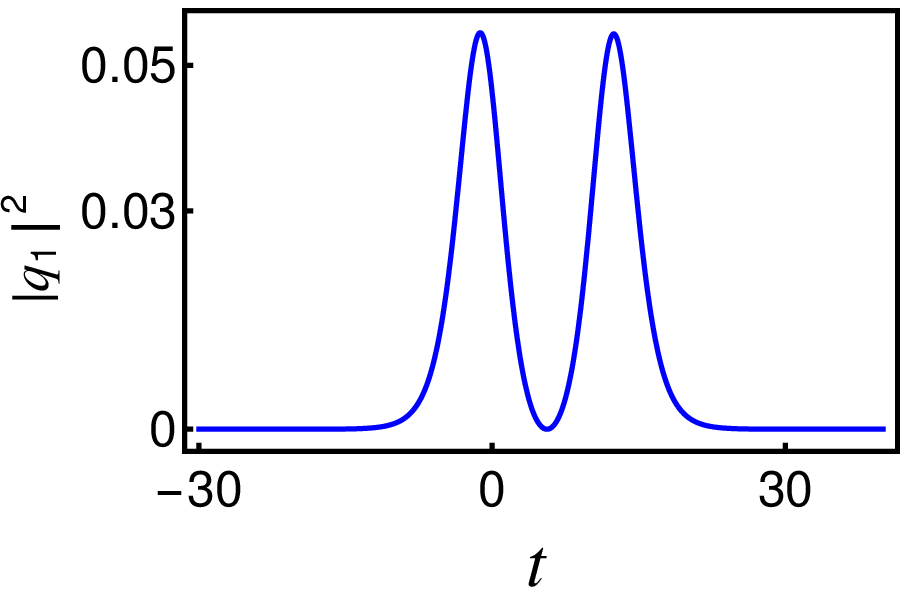}
\includegraphics[width=0.35\linewidth]{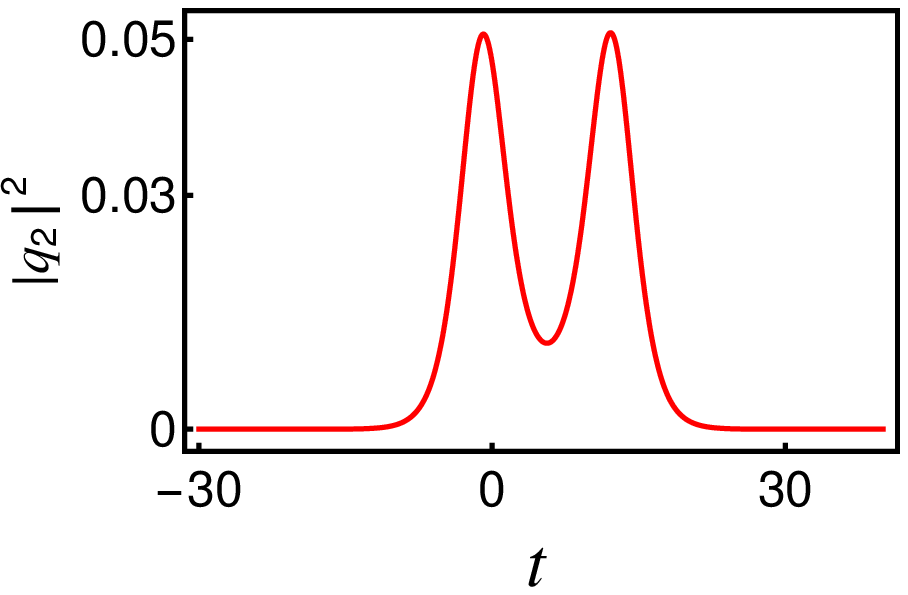}
\caption{Symmetric double hump-double hump soliton: Parameter values are $k_{1} = 0.316+ 0.5i$, $l_{1} = 0.333 + 0.5i$, $\alpha_{1}^{(1)} = 0.49 + 0.45i$ and $\alpha_{1}^{(2)} = 0.45 + 0.45i$.}
\label{f5}
\end{figure}